\pdfoutput=1

\documentclass[11pt]{article}

\usepackage[final]{acl}

\usepackage[T1]{fontenc}
\usepackage[utf8]{inputenc}

\usepackage{xcolor}
\usepackage{hyperref}
\hypersetup{
    colorlinks=true,
    linkcolor=blue,
    filecolor=magenta,
    urlcolor=cyan,
    pdftitle={PILAR: A Page-Grounded Unified Evidence Representation via an Entity-Linked Assertion Graph for Open-Domain QA Agents over Multimodal Document Corpora},
    pdfauthor={Joongmin Shin, Gyuho Shim, Jung-hun Lee, Jaehyung Seo},
}

\usepackage{times}
\usepackage{latexsym}

\usepackage{graphicx}
\usepackage{booktabs}
\usepackage{multirow}
\usepackage{adjustbox}
\usepackage{makecell}
\usepackage{colortbl}
\usepackage{enumitem}
\usepackage{placeins}
\usepackage{float}

\usepackage{amsmath}
\usepackage{amssymb}
\usepackage{algorithm}
\usepackage{algpseudocode}

\usepackage{tikz}
\usetikzlibrary{arrows.meta,positioning,fit,calc}
\usetikzlibrary{backgrounds}

\usepackage{array}
\newcolumntype{C}[1]{>{\centering\arraybackslash}p{#1}}

\definecolor{systemcolor}{HTML}{E8F1FF}
\definecolor{usercolor}{HTML}{F7FFE9}
\definecolor{assistantcolor}{HTML}{FFF7E9}
\definecolor{oursrow}{HTML}{E8F0FE}
\definecolor{bottleneckrow}{HTML}{FFF3E0}
\definecolor{deltaplus}{HTML}{0D47A1}
\definecolor{deltazero}{HTML}{9E9E9E}
\newcommand{\up}[1]{\textcolor{deltaplus}{\textbf{#1}}}

\usepackage{pgfplots}
\pgfplotsset{compat=1.18}

\usepackage[most]{tcolorbox}
\tcbuselibrary{breakable,skins}
\usepackage{microtype}

\newcommand{\ours}{\textbf{PILAR}}
\usepackage{pifont}
\newcommand{\cmark}{\textcolor{green!60!black}{\ding{51}}}
\newcommand{\xmark}{\textcolor{red!70!black}{\ding{55}}}

\title{PILAR: A Page-Grounded Unified Evidence Representation via an Entity-Linked Assertion Graph for Open-Domain QA Agents over Multimodal Document Corpora}

\author{
  Joongmin Shin\textsuperscript{1} \quad
  Gyuho Shim\textsuperscript{1} \quad
  Jung-hun Lee\textsuperscript{2*} \quad
  Jaehyung Seo\textsuperscript{3*} \\
  \textsuperscript{1}Korea University \\
  \textsuperscript{2}Korea Maritime and Ocean University \\
  \textsuperscript{3}Konkuk University \\
  \texttt{\{tlswndals13,gjshim\}@korea.ac.kr} \quad
  \texttt{it\_leejh@kmou.ac.kr} \\
  \texttt{seojae777@konkuk.ac.kr} \\
  \textsuperscript{*}Corresponding authors
}

\begin{document}
\maketitle

\begin{abstract}
Open-domain question answering (ODQA) over multimodal document corpora requires linking evidence scattered across text, tables, and figures. Existing systems often store these sources separately or retrieve only coarse pages, which weakens global evidence linking. We present \ours{}, a page-grounded unified evidence representation instantiated as an entity-linked assertion graph. \ours{} maps sentence-, table-, and figure-derived facts into a common assertion space and uses the graph as a controlled linking layer over robust page retrieval. In a shared-reader evaluation with four agent frameworks, fourteen retrieval backends, and two benchmarks, \ours{} achieves the best end-to-end EM/ANLS. Gains are largest on compositional, cross-document, and multimodal questions, with a single-shot improvement of +1.6 EM over flat retrieval, rising to +2.9 on compositional and +5.9 on 3-hop questions. Ablations show that current gains are driven mainly by the text-instantiated slice of the framework, while visual assertions help only after locality-aware filtering. We therefore position \ours{} as a unified evidence representation for multimodal ODQA rather than a standalone visual-reasoning module. Code is available at \url{https://github.com/ShinJM-maker/PILAR}.
\end{abstract}

\section{Introduction}
\label{sec:intro}

Open-domain question answering (ODQA) over multimodal document corpora is a retrieve-and-read problem in which systems must combine evidence spread across text, tables, and figures~\citep{gao2024retrievalaugmentedgenerationlargelanguage,Cho_2025_ICCV,tito2023hierarchicalmultimodaltransformersmultipage}. In the agentic setting we study, this requires more than retrieving relevant pages: the system must also connect related facts across documents and modalities while preserving the local grounding needed for reliable answer generation. As a result, the central challenge is not only page retrieval, but also how heterogeneous evidence is represented once it is retrieved.

Recent LLM-based QA agents~\citep{yao2023react,lee2024planrag,wu2023autogen} increase this need for a shared evidence representation that supports repeated query reformulation, evidence aggregation, and cross-page linking. A useful representation must preserve page-level locality for grounding, while also exposing higher-level connections among entities, assertions, and support evidence across documents. However, under matched evaluation, it remains unclear which kinds of structured evidence representations actually help multimodal ODQA agents, and whether gains come from text-derived structure, visual assertions, locality-aware ranking, or some combination of them. For brevity, we refer to the table-/figure-derived branch as visual assertions in the ablation analysis.

Existing approaches have complementary but incomplete strengths. Flat-chunk and page-level retrievers~\citep{faysse2025colpali,Cho_2025_ICCV} preserve document content, but they leave evidence as unstructured chunks or coarse pages. Text-only knowledge graphs~\citep{edge2024graphrag,guo2024lightrag} add useful entity--relation structure, but they exclude table- and figure-derived evidence from the same representational space. VLM-augmented page retrieval~\citep{jain-etal-2025-simpledoc} adds visual semantics through free-form descriptions, but these descriptions are not entity-linked or predicate-normalized, making cross-document graph traversal difficult. What is still missing is a unified representation that converts text, table, and figure evidence into the same structured assertion space, links them in a single entity-linked graph, and exposes them through a page-grounded interface for ODQA agents.

\begin{figure}[t]
\centering
\resizebox{0.98\columnwidth}{!}{%
\begin{tikzpicture}[
  font=\scriptsize,
  >=Latex,
  qbox/.style={draw, rounded corners=3pt, thick, fill=pink!10, align=center, text width=70mm, inner sep=4pt},
  entity/.style={draw, rounded corners=3pt, thick, fill=blue!8, align=center, minimum width=28mm, minimum height=8mm},
  assert/.style={draw, rounded corners=3pt, thick, fill=yellow!12, align=center, text width=30mm, inner sep=3pt},
  support/.style={draw, rounded corners=3pt, fill=green!8, align=left, text width=30mm, inner sep=3pt},
  packet/.style={draw, rounded corners=3pt, thick, fill=violet!8, align=left, text width=88mm, inner sep=4pt},
  edgelbl/.style={font=\tiny, fill=white, inner sep=1pt}
]
\node[qbox] (q) at (0,2.1) {\textbf{Query}\\At $5^\circ$C, what charging current should BatteryPack-A use?};
\node[entity] (e) at (0,0.95) {\textbf{BatteryPack-A}\\canonical entity};

\node[assert] (a1) at (-3.45,-0.35) {\textbf{Sentence assertion}\\temp $< 10^\circ$C $\Rightarrow$ cold-start mode};
\node[support] (s1) at (-3.45,-1.7) {\textbf{Support}\\Doc~1, p.3\\sentence span + section header};

\node[assert] (a2) at (0,-0.35) {\textbf{Table assertion}\\BatteryPack-A: nominal = 4A; reduced = 2A};
\node[support] (s2) at (0,-1.7) {\textbf{Support}\\Doc~1, p.3 table cell\\model-specific settings};

\node[assert] (a3) at (3.45,-0.35) {\textbf{Caption assertion}\\cold-start mode $\Rightarrow$ use reduced current};
\node[support] (s3) at (3.45,-1.7) {\textbf{Support}\\Doc~2, p.1 figure caption\\operating rule};

\node[packet] (p) at (0,-3.45) {\textbf{Page-grounded evidence packet}\\
$5^\circ$C $\Rightarrow$ cold-start mode \quad | \quad BatteryPack-A: nominal 4A / reduced 2A \quad | \quad use reduced current $\Rightarrow$ 2A};

\draw[->, thick] (q) -- (e);
\draw[->, thick] (e) -- node[edgelbl, above left]{scope} (a1.north);
\draw[->, thick] (e) -- node[edgelbl, right]{spec} (a2.north);
\draw[->, thick] (e) -- node[edgelbl, above right]{exception} (a3.north);
\draw[->, thick] (a1) -- (s1);
\draw[->, thick] (a2) -- (s2);
\draw[->, thick] (a3) -- (s3);
\draw[->, thick] (s1.south) -- ++(0,-0.25) -| (p.north west);
\draw[->, thick] (s2.south) -- (p.north);
\draw[->, thick] (s3.south) -- ++(0,-0.25) -| (p.north east);
\end{tikzpicture}%
}
\caption{Running example of \ours{}. A sentence defines when cold-start mode applies, a table lists the model-specific charging settings for BatteryPack-A, and a figure caption in a second document specifies that cold-start operation uses the reduced current. \ours{} links these supports through the canonical entity \emph{BatteryPack-A} and returns one page-grounded evidence packet that preserves both cross-document linking and local grounding.}
\label{fig:running_example}
\end{figure}
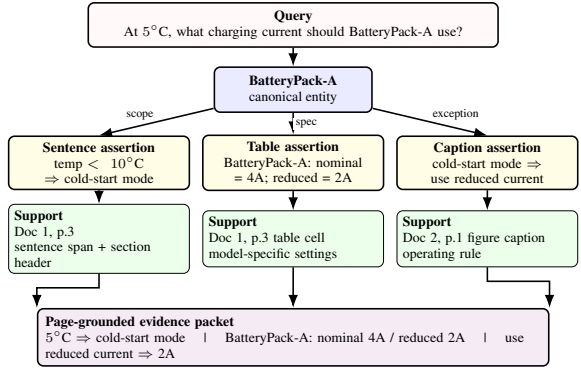

\paragraph{Running example.}
Consider a toy question: \emph{At $5^\circ$C, what charging current should BatteryPack-A use?} A sentence in Doc~1 states that \emph{cold-start mode applies below $10^\circ$C}. A table on the same page lists model-specific charging settings and gives \emph{BatteryPack-A: nominal current = 4A, reduced current = 2A}. A figure caption in Doc~2 states that \emph{in cold-start mode, the reduced current should be used to avoid cell damage}. \ours{} maps these sentence-, table-, and figure-derived facts into one page-grounded assertion neighborhood centered on the canonical entity \emph{BatteryPack-A}: the sentence grounds the operating regime, the table enumerates the candidate settings, and the caption specifies which setting applies. Instead of handing the agent three disconnected stores, \ours{} returns one page-grounded evidence packet that preserves both the local supports and the cross-document link needed to recover the correct answer, \emph{2A}. Figure~\ref{fig:running_example} visualizes this neighborhood.

To address this gap, we present \ours{}. \ours{} is a page-grounded unified evidence representation, instantiated as an entity-linked assertion graph and used as a controlled linking layer over robust page retrieval for open-domain QA agents over multimodal document corpora. \ours{} maps sentence-, table-, and figure-derived facts into a shared entity-linked assertion graph, while retaining the page-anchored support objects that license each assertion. In this design, the graph complements rather than replaces the base retriever, enabling global evidence linking across documents and modalities within a single page-grounded representation. In a shared-reader evaluation across four agent families and two benchmarks, \ours{} achieves the strongest end-to-end EM/ANLS among the compared systems, with the largest gains on compositional, cross-document, and multimodal questions. A strict matched-interface ablation further shows that the current gains are driven mainly by the text-instantiated slice of the framework, while visual assertions contribute conditionally after locality-aware filtering. We therefore view \ours{} not as a visual-assertion-first system, but as a unified multimodal evidence framework whose benefits grow as QA requires linking dispersed support while preserving local grounding.

\paragraph{Contributions.}
\begin{itemize}[leftmargin=*]
    \item \textbf{\ours{}:}
    We introduce \ours{}, a page-grounded unified evidence representation for open-domain QA agents over multimodal document corpora. \ours{} maps sentence-, table-, and figure-derived evidence into a shared structured assertion space while preserving page-anchored support and provenance.

    \item \textbf{Entity-Linked Assertion Graph as a Controlled Linking Layer:}
    We instantiate this representation as an entity-linked assertion graph and use it as a controlled linking layer over robust page retrieval. This design enables cross-document and cross-modal evidence linking without replacing a strong page retriever or sacrificing local grounding.

    \item \textbf{Strong End-to-End Gains Where Evidence Linking Matters:}
    In a shared-reader evaluation across four agent frameworks, fourteen retrieval backends, and two benchmarks, \ours{} achieves the best end-to-end EM/ANLS. In the single-shot setting, it improves EM by \textbf{+1.6} over flat retrieval overall, with larger gains of \textbf{+2.9} on compositional questions and \textbf{+5.9} on 3-hop questions.

    \item \textbf{Mechanistic Diagnosis of What Drives the Gains:}
    Matched-interface ablations show that the current gains are driven mainly by the text-instantiated slice of \ours{}, while table- and figure-derived assertions help after locality-aware filtering. This clarifies both where the framework is already effective and what bottlenecks remain.
\end{itemize}

\begin{figure*}[t]
\centering
\includegraphics[width=\textwidth]{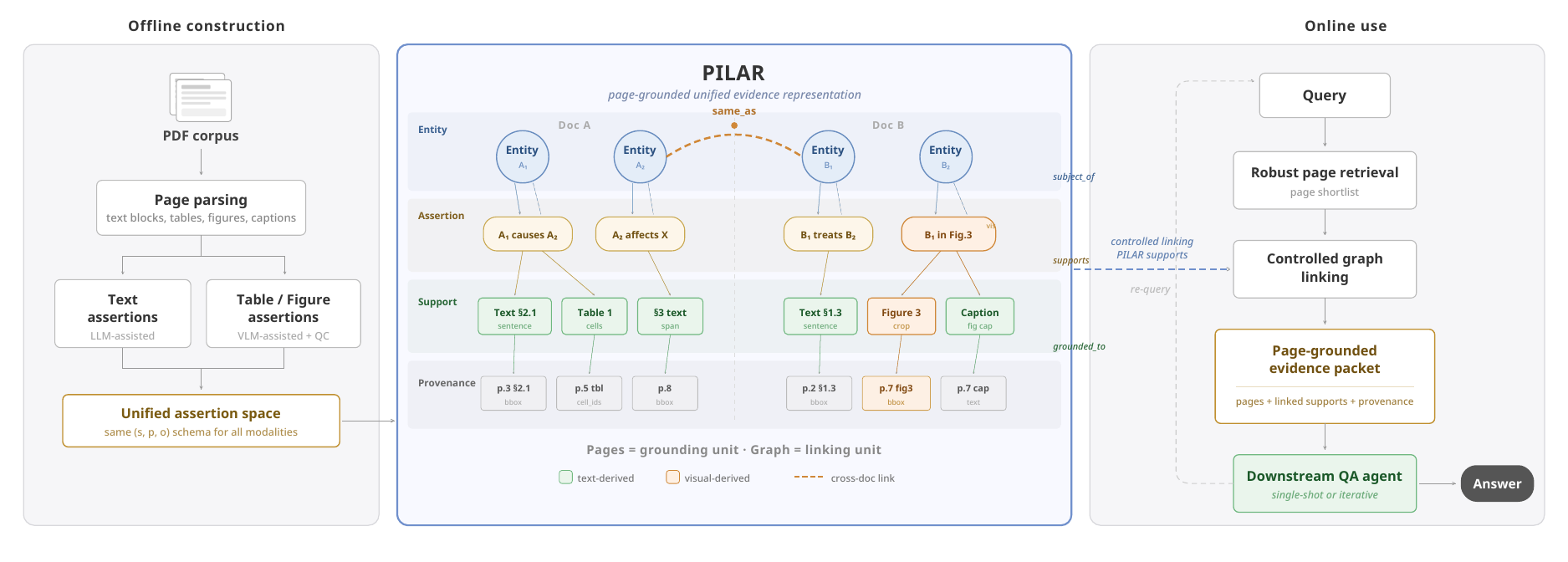}
\caption{Overview of \ours{}. Offline (left), PDFs are parsed into page-anchored units, and text-, table-, and figure-derived facts are mapped into a unified assertion space to build \ours{}, a page-grounded unified evidence representation. The graph organizes evidence into entity, assertion, support, and provenance layers, with cross-document entity bridges enabling global evidence linking while page-level provenance preserves local grounding (center). At query time (right), a robust page retriever produces candidate pages, and \ours{} is used as a controlled linking layer. The system returns a page-grounded evidence packet to a downstream QA agent, which can answer directly or issue a follow-up query.}
\label{fig:system_architecture}
\end{figure*}

\section{Related Work}

Prior work on ODQA over multimodal document corpora differs mainly in its retrieval unit and evidence representation. Chunk-based and structure-aware methods improve granularity by indexing text chunks, structural sections, or summary trees~\citep{gao2024retrievalaugmentedgenerationlargelanguage,duarte-etal-2024-lumberchunker,zhao-etal-2025-moc,yepes2024financialreportchunkingeffective,multidocfusion,sarthi2024raptorrecursiveabstractiveprocessing}. Text-centric graph methods add global structure through chunk-link or entity--relation graphs, but they remain primarily text-derived~\citep{liu-etal-2025-hoprag,edge2024graphrag,guo2024lightrag}. Page-level multimodal retrievers instead preserve local visual grounding by retrieving whole pages or page-image representations, sometimes with VLM-generated page descriptions or page-graph expansion~\citep{Cho_2025_ICCV,vdocrag2025,faysse2025colpali,jain-etal-2025-simpledoc,wu-etal-2025-molorag}. These families provide complementary strengths, but each leaves part of the representation problem unresolved; Appendix~\ref{sec:rw_appendix} sharpens this positioning with a method-level summary.

The common limitation is that text, table, and figure evidence are usually not expressed in the same structured space. Chunk-based methods remain locally fragmented, page-based multimodal methods stay coarse under limited context budgets, and text-only graph methods exclude non-text evidence from the same evidence object. \ours{} is designed to fill this gap: it is a page-grounded unified evidence representation, instantiated as an entity-linked assertion graph, that links sentence-, table-, and figure-derived evidence while preserving page-level provenance for downstream QA agents.

Table~\ref{tab:positioning_main} summarizes the central design claim of \ours{}: unlike prior families, it unifies text-, table-, and figure-derived evidence in the same assertion space while preserving page-grounded support and using graph structure only as a controlled linking layer over robust page retrieval.

\begin{table}[t]
\centering
\small
\renewcommand{\arraystretch}{1.12}
\setlength{\tabcolsep}{3pt}
\begin{tabular*}{\columnwidth}{@{\extracolsep{\fill}}lccc}
\toprule
\makecell[l]{\textbf{Method}\\\textbf{family}} &
\makecell{\textbf{Shared}\\\textbf{space}} &
\makecell{\textbf{Page}\\\textbf{grounding}} &
\makecell{\textbf{Linking}\\\textbf{layer}} \\
\midrule
\makecell[l]{Chunk /\\structure-aware} & \xmark & \xmark & \xmark \\
\makecell[l]{Text-centric\\graph} & \xmark & \xmark & \cmark \\
\makecell[l]{Page-level\\multimodal} & \xmark & \cmark & \xmark \\
\textbf{\ours{}} & \textbf{\cmark} & \textbf{\cmark} & \textbf{\cmark} \\
\bottomrule
\end{tabular*}
\caption{\textbf{Why \ours{} is different.} Prior families satisfy only part of the design goal: chunk methods improve granularity, text-centric graph methods add global structure, and page-level multimodal methods preserve local grounding. \ours{} combines all three in one page-grounded unified evidence representation.}
\label{tab:positioning_main}
\end{table}

\paragraph{Design takeaway.} \ours{} is built around a simple separation of roles: pages are the grounding unit, while the entity-linked assertion graph is the linking unit. This lets the system connect sentence-, table-, and figure-derived evidence across documents without replacing a strong page retriever or losing local support. The empirical pattern in our results matches this design: gains are concentrated on compositional, cross-document, and multimodal questions, while the main remaining bottlenecks lie in base page recall and reader capacity.

\section{PILAR: Page-grounded Integrated Linked-Assertion Representation}
\label{sec:method}

We present \ours{} (\emph{Page-grounded Integrated Linked-Assertion Representation}), a page-grounded unified evidence representation via an entity-linked assertion graph for open-domain QA agents over multimodal document corpora. The key idea is to convert sentence-, table-, and figure-derived evidence into the same structured assertion space, so that heterogeneous evidence can be linked through shared entities and normalized predicates rather than stored in separate modality-specific indexes. At the same time, each assertion remains tied to page-grounded support, including the original text span, table region, figure crop, or caption that licenses it. \ours{} is agent-compatible: it exposes the same page-grounded evidence packet interface to downstream QA agents, while planning and action policies remain outside the framework. This design lets \ours{} support global evidence linking across documents and modalities while preserving the local grounding needed for reliable answer generation.

\subsection{System Overview}

\ours{} follows a four-stage pipeline that separates evidence extraction, graph construction, retrieval, and answer packaging. First, each PDF is parsed into page-anchored units such as text blocks, tables, figures, and captions, and a lightweight hierarchy parser assigns each unit a section path. Second, the system converts these heterogeneous units into a unified entity-linked assertion graph in which text-, table-, and figure-derived facts share the same schema and can be connected through canonical entities. Third, at query time, a standard page retriever produces an initial shortlist, and the graph is used as a controlled linking layer for cross-document and cross-modal evidence linking. Finally, the system assembles a compact page-grounded evidence packet and passes it to the reader.

From the agent's perspective, \ours{} acts as a retrieval-and-packaging backend. An agent issues a question or follow-up query, receives a page-grounded evidence packet rather than raw graph neighborhoods or free-form summaries, and then either answers from that packet or issues another query. The same interface is reused across single-shot prompting, iterative query reformulation, and planner--executor loops; only the agent's planning policy changes, while \ours{} continues to provide the same page-grounded evidence object.

A key design choice is to combine page-grounded retrieval with assertion-level evidence linking. Pages preserve the local multimodal context needed to interpret nearby paragraphs, tables, figures, and captions together, while the assertion graph provides the finer semantic structure needed to connect related evidence across page and document boundaries. In this sense, \ours{} does not replace robust page retrieval with graph traversal; it uses the graph as a controlled linking layer to unify heterogeneous evidence on top of a stable page-grounded base.

\subsection{Shared Assertion Space for Text, Table, and Figure Evidence}
\label{sec:vlm_extraction}

\ours{} converts text, table, and figure content into the same structured assertion space. Text blocks are converted into subject--predicate--object assertions with optional qualifiers. Tables, charts, diagrams, and figures are processed with a VLM over cropped regions together with OCR and caption text, and the VLM outputs assertions in the same schema.

This shared assertion space is a key difference from description-based multimodal retrieval. Rather than storing free-form table or figure summaries, \ours{} turns table- and figure-derived evidence into structured, entity-linked claims that can be merged with text-derived claims and traversed by the same retriever. A sentence, a table value, and a figure-derived fact about the same entity therefore become neighbors in one graph instead of isolated items in separate modality-specific stores.

Because table- and figure-derived extraction is noisier than text extraction, we apply lightweight quality control before adding these assertions to the graph. We keep only assertions that are grounded in the corresponding table or figure crop and verify key values against OCR or nearby text when available. Detailed prompting, normalization, and filtering rules are deferred to Appendix~\ref{sec:method_appendix}.

\subsection{Unified Entity-Linked Assertion Graph}

On top of this shared assertion space, \ours{} organizes corpus evidence into four connected layers. The \textbf{entity layer} stores canonical entities and cross-document aliases. The \textbf{assertion layer} stores normalized claims in subject--predicate--object form. The \textbf{support layer} stores the evidence units that justify these claims, including text spans, tables, figure regions, and captions. The \textbf{provenance layer} records where each support comes from, such as document ID, page number, section path, and bounding box.

This design separates two roles that are often conflated in document retrieval. The entity and assertion layers support \emph{global evidence linking}: they connect semantically related evidence across the corpus. The support and provenance layers provide \emph{local grounding}: they preserve the exact page context needed to verify and interpret a claim. As a result, \ours{} can follow cross-document semantic links without losing page-level interpretability.

\subsection{Page Retrieval with Controlled Graph Expansion}
\label{sec:retrieval}

At query time, \ours{} begins with a standard hybrid page retriever:
\[
\begin{aligned}
s_{\text{base}}(p,q)
&= 0.4\,\mathrm{BM25}_{\text{norm}}(p,q) \\
&\quad + 0.6\,\mathrm{dense}(p,q),
\end{aligned}
\]
where $q$ is the query and $p$ is a candidate page. This stage provides an initial page shortlist with strong lexical and semantic coverage.

The graph then acts as a \emph{controlled linking layer} on top of this shortlist rather than replacing the base retriever. Starting from high-confidence seed pages, \ours{} applies three lightweight signals. First, it favors \emph{local continuity} through adjacent-page and same-section links, which helps when evidence spills across page boundaries. Second, it performs \emph{alias expansion} so that alternative names of the same entity can recover supporting pages from other documents. Third, it performs \emph{graph expansion} through entity--assertion--support links to recover pages that are semantically related to the query but may not be strong lexical matches on their own. A relevance gate suppresses weak expansions and prevents graph drift.

Built on this design, the base retriever provides robust corpus-scale recall, while the graph improves evidence connectivity inside the candidate set. In practice, this lets \ours{} recover cross-document and cross-modal support without sacrificing page-level stability.

\subsection{Page-Grounded Evidence Packet Construction}

The final step assembles a reader packet under a fixed token budget. Instead of concatenating raw pages in score order, \ours{} packages \emph{page-anchored evidence neighborhoods}: the selected page context together with the supporting text or visual units on that page and the minimal local scope needed to interpret them correctly, such as captions or nearby headers.

This design keeps the input readable for the reader while reducing irrelevant context. Importantly, the packet remains page-grounded even when retrieval uses graph links. The graph supplies the semantic connections across pages and documents, while the packet preserves the local evidence needed for verification. Operationally, this packet is the object consumed by downstream QA agents: after inspecting it, an agent may answer directly or issue a refined follow-up query, in which case \ours{} reruns the same retrieve--link--pack procedure and returns another packet under the same interface. Detailed scoring, thresholds, and serialization rules are provided in Appendix~\ref{sec:retrieval_appendix} and Appendix~\ref{sec:packing_details}.
\section{Experiments}
\label{sec:experimental}

\begin{table*}[t]
\centering
\scriptsize
\renewcommand{\arraystretch}{1.05}
\setlength{\tabcolsep}{2.0pt}
\begin{adjustbox}{width=\textwidth, keepaspectratio}
\begin{tabular}{l cccc cccc}
\toprule
 & \multicolumn{4}{c}{M3DocVQA (EM / ANLS)}
 & \multicolumn{4}{c}{Frames (EM / ANLS)} \\
\cmidrule(lr){2-5}\cmidrule(lr){6-9}
Backend &
Naive RAG & ReAct & PlanRAG & AutoGen &
Naive RAG & ReAct & PlanRAG & AutoGen \\
\midrule
\multicolumn{9}{l}{\textit{Text chunk-based RAG}} \\
\quad Flat chunk
& 32.0\,/\,34.4 & 31.4\,/\,34.1 & 27.4\,/\,29.5 & 31.8\,/\,34.5
& 13.8\,/\,16.7 & 11.6\,/\,14.5 & 9.1\,/\,11.2 & 13.4\,/\,16.6 \\
\quad LumberChunker
& 31.2\,/\,33.6 & 30.0\,/\,32.6 & 26.4\,/\,28.2 & 31.0\,/\,33.4
& 13.2\,/\,16.1 & 11.4\,/\,13.8 & 9.0\,/\,11.2 & 13.0\,/\,15.8 \\
\quad Meta Chunker
& 31.0\,/\,33.4 & 29.8\,/\,32.4 & 26.4\,/\,28.0 & 31.0\,/\,33.4
& 13.0\,/\,15.9 & 11.2\,/\,13.6 & 8.8\,/\,11.0 & 12.8\,/\,15.6 \\
\quad Structural chunking
& 30.4\,/\,32.8 & 29.2\,/\,31.8 & 25.8\,/\,27.6 & 30.2\,/\,32.6
& 12.6\,/\,15.4 & 10.8\,/\,13.4 & 8.6\,/\,10.8 & 12.4\,/\,15.0 \\
\midrule
\multicolumn{9}{l}{\textit{Hierarchical / structure-aware retrieval}} \\
\quad RAPTOR
& 31.4\,/\,33.8 & 30.2\,/\,32.8 & 26.6\,/\,28.4 & 31.2\,/\,33.6
& 13.0\,/\,15.9 & 11.2\,/\,13.8 & 8.8\,/\,11.0 & 12.8\,/\,15.6 \\
\quad MultiDocFusion
& 31.4\,/\,33.9 & 30.0\,/\,32.6 & 27.2\,/\,29.1 & 32.0\,/\,33.7
& 12.2\,/\,15.6 & 10.8\,/\,13.3 & 8.3\,/\,11.0 & 11.6\,/\,14.3 \\
\midrule
\multicolumn{9}{l}{\textit{Graph-augmented chunk retrieval}} \\
\quad HopRAG
& 31.6\,/\,34.0 & 30.4\,/\,33.0 & 26.8\,/\,28.6 & 31.4\,/\,33.8
& 13.2\,/\,16.1 & 11.4\,/\,14.0 & 9.0\,/\,11.2 & 13.0\,/\,15.8 \\
\midrule
\multicolumn{9}{l}{\textit{Entity-KG GraphRAG}} \\
\quad MS GraphRAG
& 32.2\,/\,34.6 & 29.8\,/\,32.5 & 26.6\,/\,28.6 & 31.2\,/\,34.0
& 13.6\,/\,16.3 & 11.2\,/\,13.8 & 9.3\,/\,11.1 & 13.0\,/\,16.0 \\
\quad LightRAG
& 31.6\,/\,34.2 & 31.0\,/\,33.5 & 26.4\,/\,28.5 & 31.6\,/\,34.3
& 12.6\,/\,15.5 & 11.4\,/\,13.9 & 9.1\,/\,11.4 & 13.6\,/\,16.0 \\
\midrule
\multicolumn{9}{l}{\textit{Page-level multimodal RAG}} \\
\quad M3DocRAG
& 19.0\,/\,25.2 & 16.0\,/\,16.6 & 10.0\,/\,12.7 & 19.0\,/\,25.2
& 12.0\,/\,16.5 & 4.0\,/\,5.3 & 1.0\,/\,3.2 & 11.0\,/\,15.2 \\
\quad VDocRAG
& 20.4\,/\,26.6 & 17.2\,/\,17.6 & 10.8\,/\,13.4 & 20.4\,/\,26.6
& 12.4\,/\,17.1 & 4.2\,/\,5.4 & 1.0\,/\,3.4 & 11.4\,/\,15.6 \\
\midrule
\multicolumn{9}{l}{\textit{VLM-augmented page retrieval}} \\
\quad SimpleDoc
& 24.0\,/\,28.7 & 29.0\,/\,33.2 & 20.0\,/\,22.9 & 24.0\,/\,29.6
& 13.0\,/\,19.5 & 7.0\,/\,14.0 & 6.0\,/\,10.7 & 11.0\,/\,18.0 \\
\midrule
\multicolumn{9}{l}{\textit{Multimodal GraphRAG}} \\
\quad MoLoRAG
& 19.4\,/\,25.6 & 16.4\,/\,17.0 & 10.4\,/\,13.0 & 19.4\,/\,25.6
& 12.2\,/\,16.8 & 4.2\,/\,5.5 & 1.2\,/\,3.4 & 11.2\,/\,15.4 \\
\midrule
\multicolumn{9}{l}{\textit{Page-grounded unified evidence representation (ours)}} \\
\rowcolor{oursrow}
\quad \textbf{\ours{}}
& \textbf{33.6\,/\,36.3} & \textbf{31.4\,/\,34.5} & \textbf{28.0\,/\,29.8} & \textbf{32.6\,/\,35.3}
& \textbf{15.0\,/\,17.9} & \textbf{12.4\,/\,15.2} & \textbf{10.5\,/\,12.5} & \textbf{14.5\,/\,17.5} \\
\bottomrule
\end{tabular}
\end{adjustbox}
\caption{\textbf{Main comparison.} Fourteen backends $\times$ four agents; each cell shows EM\,/\,ANLS. All text-based and graph backends share the same BM25+dense page retriever and reader (Qwen3-VL-8B~\citep{qwen3vl}); page-based multimodal methods retain their native retrieval and packing pipeline under the same reader and evaluation protocol. \ours{} yields the strongest overall profile in this comparison, with the clearest gains on M3DocVQA and competitive performance on Frames; the Naive RAG gain is significant on M3DocVQA ($p$\,=\,0.048; paired approximate randomization~\citep{yeh-2000-accurate}), and pooling across agents remains significant ($p$\,=\,0.019). Additional ROUGE-L results and oracle analyses appear in Appendix~\ref{sec:bottleneck_appendix}.}
\label{tab:main_results}
\end{table*}

We evaluate \ours{} through four complementary analyses: a main comparison across agent families and retrieval backends, query-attribute localization, a strict matched-interface ablation, and bottleneck-oriented diagnostics. Appendix~\ref{sec:bottleneck_appendix} provides extended analyses on pruning, reader-family sensitivity, visual-source audits, and modality oracles. Unless noted otherwise, all systems share the same reader (Qwen3-VL-8B~\citep{qwen3vl}) and the same evaluation protocol.

\subsection{Setup}
\label{sec:exp}

To ensure a fair comparison, all baselines use the same reader (Qwen3-VL-8B~\citep{qwen3vl}), prompt template, and core evaluation protocol. For text-based and graph-oriented backends, we use a strict matched-interface setting in which all methods share the same BM25+dense base page retriever under one token-budgeted evaluation framework; page-based multimodal baselines keep their native retrieval and packaging pipelines because those components are central to their design. We evaluate on M3DocVQA~\citep{Cho_2025_ICCV} (2,441 questions) and Frames~\citep{krishna2025frames} (824 questions), reporting EM and ANLS. The four agent frameworks are Naive RAG, ReAct~\citep{yao2023react}, PlanRAG~\citep{lee2024planrag}, and AutoGen~\citep{wu2023autogen}. Across these agents, we compare fourteen retrieval backends spanning text chunking, hierarchical retrieval, graph-augmented chunk retrieval, entity-KG methods, page-level multimodal retrieval, VLM-augmented page retrieval, multimodal page graphs, and \ours{}; full configurations appear in Appendix~\ref{sec:agent_baselines}, Appendix~\ref{sec:kg_baselines}, and Appendix~\ref{sec:retrieval_appendix}.

\subsection{Main Comparison: Agent $\times$ Retrieval Backend}
\label{sec:main_results}

Table~\ref{tab:main_results} shows the main comparison across four agent families under the shared-reader setup. In this controlled setting, \ours{} gives the best results on all M3DocVQA settings and on all Frames EM settings, while remaining competitive on Frames ANLS. The clearest gains appear on M3DocVQA. For Naive RAG on M3DocVQA, the gain is statistically significant ($\Delta$EM\,=\,+1.6, $p$\,=\,0.048; $\Delta$ANLS\,=\,+1.9, $p$\,=\,0.028), and pooling across all agents is also significant ($p$\,=\,0.019). On Frames, no individual comparison reaches the significance threshold, so we treat it as a directional support-chain stress test rather than conclusive evidence.

Average gains alone understate the value of the method. Table~\ref{tab:attribute_summary} shows that the improvements are concentrated on compositional, two-document, and multimodal questions---the cases where answer support is least likely to stay on one page and most likely to require cross-page or cross-document linking. This is the setting where a page-grounded unified evidence representation should help most. Table~\ref{tab:strict_ablation_main} then explains this pattern: under the current setup, most of the measurable gain comes from the text-instantiated slice of \ours{}, while visual assertions help only when noisy expansion paths are controlled.

\subsection{Query-Attribute Localization}
\label{sec:attribute_main}

Table~\ref{tab:attribute_summary} identifies the query dimensions where \ours{} helps most. The key result is not just a higher average EM, but that the gains appear exactly where an entity-linked retrieval index should matter most. Relative to flat chunk retrieval, the gain grows from +1.4 EM on 1-hop queries to +1.9 on 2-hop and +5.9 on the sampled 3-hop subset, and from +1.1 on single-document to +2.9 on two-document questions. Even when measured against the strongest non-\ours{} baseline per subset (Table~\ref{tab:attribute_summary}, $\Delta_{\text{best}}$ column), gains remain positive on all compositional, cross-document, and multimodal subsets.

This distribution supports our main methodological claim. \ours{} is not meant to improve every question equally. It helps most when answer support must be assembled across pages, documents, or modalities rather than retrieved from one locally matched page. Because the 1-hop and Simple buckets almost fully overlap in M3DocVQA, they should be viewed as one easy-query stratum rather than as independent evidence. The sampled 3-hop bucket shows the largest gain, but its small size means that this magnitude should be treated as suggestive rather than definitive.

\begin{table*}[t]
\centering
\small
\setlength{\tabcolsep}{4pt}
\renewcommand{\arraystretch}{1.10}
\begin{tabular}{@{}l c c c c c @{\hskip 14pt} l c c c c c@{}}
\toprule
\multicolumn{6}{c}{\textbf{By Hop Count \& Question Type}} & \multicolumn{6}{c}{\textbf{By Document Count \& Modality}} \\
\cmidrule(r){1-6}\cmidrule(l){7-12}
Subset & \% & \ours{} & Flat & Best-other & $\Delta_{\text{best}}$
& Subset & \% & \ours{} & Flat & Best-other & $\Delta_{\text{best}}$ \\
\midrule
1-hop & 61.8 & 42.2 & 40.8 & 41.4\textsuperscript{G} & \up{+0.8}
& 1 document & 34.6 & 40.9 & 39.8 & 40.4\textsuperscript{G} & \up{+0.5} \\
2-hop & 31.3 & 16.1 & 14.2 & 14.9\textsuperscript{G} & \up{+1.2}
& 2 documents & 56.8 & 31.7 & 28.8 & 29.5\textsuperscript{G} & \up{+2.2} \\
3-hop$^{\dagger}$ & 6.9 & 41.2 & 35.3 & 35.9\textsuperscript{G} & \up{+5.3}
& & & & & & \\
\midrule
Simple & 61.8 & 42.2 & 40.8 & 41.4\textsuperscript{G} & \up{+0.8}
& Text-only & 31.1 & 59.1 & 57.1 & 57.8\textsuperscript{L} & \up{+1.3} \\
Compose & 28.1 & 15.1 & 12.2 & 12.2\textsuperscript{F} & \up{+2.9}
& Multi-modal & 39.2 & 20.1 & 18.0 & 18.7\textsuperscript{G} & \up{+1.4} \\
Compare$^{\dagger}$ & 7.9 & 35.9 & 33.3 & 33.3\textsuperscript{F} & \up{+2.6}
& & & & & & \\
\bottomrule
\end{tabular}
\caption{\textbf{Query-attribute localization} (EM, Naive RAG, M3DocVQA). $\Delta_{\text{best}}$ = \ours{} $-$ best non-\ours{} baseline. Superscripts: \textsuperscript{G}\,MS GraphRAG, \textsuperscript{L}\,LightRAG, \textsuperscript{F}\,Flat chunk. Even against the strongest competitor in each subset, \ours{} shows consistent gains on compositional (+2.9), two-document (+2.2), and multi-modal (+1.4) questions. This pattern matches the intended role of an entity-linked graph, which should help most when answer support must be linked across multiple pages or documents rather than retrieved locally. $^{\dagger}$Small buckets; interpret cautiously. Percentages may not sum to 100\% (minor categories omitted).}
\label{tab:attribute_summary}
\end{table*}

\subsection{Strict Matched-Interface Ablation}
\label{sec:strict_ablation_main}

Table~\ref{tab:strict_ablation_main} reports the main matched-interface ablation in the single-shot M3DocVQA setting. All rows use the same reader, prompt, packet budget, and evaluation split; only the retrieval architecture and lightweight reranking components change. The largest single gain comes from the text-only slice of \ours{} (+1.2 EM over page-only retrieval). Visual assertions are slightly harmful when added without locality-aware filtering (-0.4 EM). Adding the locality prior before the text-only slice also lowers EM slightly (33.2 $\rightarrow$ 33.0), because the adjacent-page bonus can push a more relevant graph-retrieved page out of the fixed-budget packet. This shows that locality is a noisy prior whose value depends on what it interacts with. When paired with visual assertions, however, the same prior constrains noisy visual expansion paths, recovers +0.4 EM (33.0 $\rightarrow$ 33.4), and helps the full system reach the best overall score (33.6). Paired significance testing shows that the full pipeline significantly outperforms page-only retrieval ($\Delta$EM\,=\,+1.6, $p$\,=\,0.048; $\Delta$ANLS\,=\,+1.9, $p$\,=\,0.028). Because the intermediate ablation variants are not individually significant, the components appear to interact synergistically rather than contributing purely additive gains (full table in Appendix~\ref{sec:significance_supp}). Overall, we interpret \ours{} primarily as a page-grounded unified evidence representation whose current gains are driven by its text-instantiated slice, with visual assertions acting as conditional entity-linking cues that help only when locality-aware filtering removes noisy expansions.

\begin{table}[t]
\centering
\small
\setlength{\tabcolsep}{3pt}
\renewcommand{\arraystretch}{1.12}
\resizebox{\columnwidth}{!}{%
\begin{tabular}{@{}lcccc cccc@{}}
\toprule
& \multicolumn{4}{c}{Components} & & \multicolumn{3}{c}{Answer Quality} \\
\cmidrule(lr){2-5}\cmidrule(lr){7-9}
Variant & Local & Text slice & Visual & Gate & & EM & ANLS & ROUGE-L \\
\midrule
\multicolumn{9}{@{}l}{\textit{\footnotesize Page-level baselines}} \\[1pt]
Page-only retrieval & \xmark & \xmark & \xmark & \xmark & & 32.0 & 34.4 & 37.3 \\
Page-only + local prior & \cmark & \xmark & \xmark & \xmark & & 32.4 & 35.6 & 38.6 \\
\addlinespace[3pt]
\multicolumn{9}{@{}l}{\textit{\footnotesize + PILAR text-only slice}} \\[1pt]
PILAR (text-only slice) & \xmark & \cmark & \xmark & \xmark & & 33.2 & 36.1 & 38.9 \\
PILAR (text-only) + local & \cmark & \cmark & \xmark & \xmark & & 33.0 & 36.1 & 39.0 \\
\addlinespace[3pt]
\multicolumn{9}{@{}l}{\textit{\footnotesize + Visual assertions}} \\[1pt]
PILAR (text+visual slice) & \xmark & \cmark & \cmark & \xmark & & 32.8 & 35.6 & 38.2 \\
PILAR (text+visual) + local & \cmark & \cmark & \cmark & \xmark & & 33.4 & 36.1 & 38.9 \\
\midrule
\rowcolor{oursrow}
\textbf{Final} (+ gate) & \cmark & \cmark & \cmark & \cmark & & \textbf{33.6} & \textbf{36.3} & \textbf{39.1} \\
\bottomrule
\end{tabular}}
\caption{\textbf{Strict matched-interface ablation.} All rows share the same reader, prompt, packet budget, and evaluation split. The text-only slice of \ours{} provides the largest single gain (+1.2 EM over page-only). Visual assertions hurt in isolation but recover +0.4 EM once locality-aware filtering is added.}
\label{tab:strict_ablation_main}
\end{table}

\subsection{Bottleneck-Oriented Diagnostics}
\label{sec:bottleneck_main}

Gold-page oracle analysis shows substantial remaining headroom: replacing retrieved context with gold-page text raises EM from 33.6 to 42.4 (+8.8~pp), while adding adjacent pages gives no further benefit. Stage-loss controls point to the same bottleneck: base recall dominates the remaining error (R0+R1=23.4\%), reranking contributes additional loss (R2=12.6\%), and packet serialization adds essentially none (R3=0). We therefore interpret locality and graph priors as evidence-ranking signals rather than guarantees of gold-page preservation. Full oracle tables, stage-loss analyses, and modality-specific diagnostics appear in Appendix~\ref{sec:bottleneck_appendix}.

Taken together, the main comparison, query-attribute analysis, and strict ablations support a precise interpretation of \ours{}. Its main contribution is a page-grounded unified evidence representation, not a claim that visual assertions alone already drive the gains. Under the current setup, most measurable improvement comes from the text-instantiated slice of the framework, while visual assertions are noisy in isolation and become helpful only after locality-aware filtering. The strongest gains therefore appear exactly where cross-document or cross-modal evidence must be linked under one shared assertion space.

\section{Conclusion}

\label{sec:conclusion}

\ours{} is a page-grounded unified evidence representation, instantiated as an entity-linked assertion graph and used as a controlled linking layer over robust page retrieval for open-domain QA agents over multimodal document corpora. The central idea is to map sentence-, table-, and figure-derived evidence into the same entity-linked assertion space, so that cross-document and cross-modal support can be linked within one page-grounded representation without sacrificing local grounding. Empirically, \ours{} delivers the strongest end-to-end performance in our shared-reader setting, with the clearest gains on compositional, cross-document, and multimodal questions, where evidence linking rather than shallow local matching is the main difficulty. At the same time, our ablations show that the current improvements are driven mainly by the text-instantiated slice of the framework, while visual assertions remain conditionally useful after locality-aware filtering. We therefore view the main contribution of this paper as a page-grounded unified evidence representation whose present gains come from a strong text-derived backbone, and whose longer-term promise is to absorb richer visual assertions into the same representation. The main remaining bottlenecks are base page recall, support-chain recovery, and reader capacity, suggesting that the next gains will come less from adding graph complexity than from improving the retriever and reader around this page-grounded unified evidence representation.

\section*{Limitations}
\label{sec:limitations}

Several subgroup analyses use relatively small query buckets, so they should be interpreted with caution. On M3DocVQA, the full pipeline improves significantly over page-only retrieval ($p$\,=\,0.048), but the intermediate ablation variants are not individually significant. We therefore view the component-level decomposition as informative but not fully confirmatory on its own. The small and interaction-dependent gains from visual assertions let us reject a visual-assertion-first interpretation, but they do not show how widely this interaction pattern will generalize to other datasets or new agent architectures. On Frames, no individual comparison is statistically significant, likely because of limited power ($n$\,=\,824, base EM\,$\approx$\,10\%). In addition, although our backend transfers across agent frameworks at the interface level, the best retrieval hyperparameters are less stable; settings tuned for single-shot use do not always transfer well to multi-step agents. Our modality-specific conclusions are also tied to the tested pipeline and model family: at the 8B scale, OCR text clearly outperforms raw page images, although stronger and better-tuned VLMs may change this result. Finally, the Frames stress test shows that our current objective favors evidence aggregation more than strict reconstruction of long support chains. Further gains in that regime may require a different retrieval objective rather than more tuning of the current retrieval stack.

\section*{Acknowledgments}

This work was partly supported by Institute of Information \& communications Technology Planning \& Evaluation (IITP, AI Computing Support Project for R\&D) grant funded by the Korea government(MSIT) (High-Performance Research AI Computing Infrastructure Support at the 2 PFLOPS Scale, Project Number: RS-2026-25505492, Contribution Rate: 25\%) and This research was supported by Culture, Sports and Tourism R\&D Program, funded by the Ministry of Culture, Sports and Tourism, through the Korea Culture Technology Planning and Evaluation Institute, an affilated institute of the Korea Creative Content Agency grant in 2026 (Project Name: Development of an AI Agent Integrating Korean Language Knowledge for Personalized Language Consultation Services, Project Number: RS-2026-25506607, Contribution Rate: 25\%) and This work was supported by the Institute of Information \& Communications Technology Planning \& Evaluation (IITP) grant funded by the Korea government (MSIT) (IITP-2026-RS-2026-25615817, AI Star Fellowship Support Program) and This work was supported by the National IT Industry Promotion Agency (NIPA) grant funded by the Korea government (MSIT) (S1301-26-1007, Development of AI-Based Advanced Static Analysis and Automatic Generation of Software Quality Reports for Weapon Systems).

% bibliographystyle already set by acl.sty
\bibliography{custom}

\begin{thebibliography}{27}
\providecommand{\natexlab}[1]{#1}

\bibitem[{Bai et~al.(2025)}]{qwen3vl}
Shuai Bai et~al. 2025.
\newblock \href {https://arxiv.org/abs/2511.21631} {{Qwen3-VL} technical
  report}.
\newblock \emph{arXiv preprint arXiv:2511.21631}.

\bibitem[{Banerjee and Lavie(2005)}]{banerjee2005meteor}
Satanjeev Banerjee and Alon Lavie. 2005.
\newblock {METEOR}: An automatic metric for {MT} evaluation with improved
  correlation with human judgments.
\newblock In \emph{Proceedings of the ACL Workshop on Intrinsic and Extrinsic
  Evaluation Measures for Machine Translation and/or Summarization}, pages
  65--72.

\bibitem[{Biten et~al.(2019)Biten, Tito, Mafla, Gomez, Rusinol, Valveny,
  Jawahar, and Karatzas}]{biten2019scene}
Ali~Furkan Biten, Ruben Tito, Andres Mafla, Lluis Gomez, Marcal Rusinol, Ernest
  Valveny, C.~V. Jawahar, and Dimosthenis Karatzas. 2019.
\newblock Scene text visual question answering.
\newblock In \emph{Proceedings of the IEEE/CVF International Conference on
  Computer Vision (ICCV)}, pages 4291--4301.

\bibitem[{Cho et~al.(2025)Cho, Mahata, Irsoy, He, and Bansal}]{Cho_2025_ICCV}
Jaemin Cho, Debanjan Mahata, Ozan Irsoy, Yujie He, and Mohit Bansal. 2025.
\newblock \href {https://doi.org/10.1109/ICCVW69036.2025.00649} {{M3DocVQA}:
  Multi-modal multi-page multi-document understanding}.
\newblock In \emph{Proceedings of the IEEE/CVF International Conference on
  Computer Vision (ICCV) Workshops}, pages 6178--6188.

\bibitem[{Duarte et~al.(2024)Duarte, Marques, Gra{\c{c}}a, Freire, Li, and
  Oliveira}]{duarte-etal-2024-lumberchunker}
Andr{\'e}~V. Duarte, Jo{\~a}o Marques, Miguel Gra{\c{c}}a, Miguel Freire, Lei
  Li, and Arlindo~L. Oliveira. 2024.
\newblock \href {https://arxiv.org/abs/2406.17526} {{LumberChunker}: Long-form
  narrative document segmentation}.
\newblock \emph{arXiv preprint arXiv:2406.17526}.

\bibitem[{Edge et~al.(2024)Edge, Trinh, Cheng, Bradley, Chao, Mody, Truitt, and
  Larson}]{edge2024graphrag}
Darren Edge, Ha~Trinh, Newman Cheng, Joshua Bradley, Alex Chao, Apurva Mody,
  Steven Truitt, and Jonathan Larson. 2024.
\newblock \href {https://arxiv.org/abs/2404.16130} {From local to global: A
  graph {RAG} approach to query-focused summarization}.
\newblock \emph{Preprint}, arXiv:2404.16130.

\bibitem[{Faysse et~al.(2025)Faysse, Sibille, Wu, Omrani, Viaud, Hudelot, and
  Colombo}]{faysse2025colpali}
Manuel Faysse, Hugues Sibille, Tony Wu, Bilel Omrani, Gautier Viaud, C{\'e}line
  Hudelot, and Pierre Colombo. 2025.
\newblock \href {https://openreview.net/forum?id=ogjBpZ8uSi} {{ColPali}:
  Efficient document retrieval with vision language models}.
\newblock In \emph{The Thirteenth International Conference on Learning
  Representations}. OpenReview.net.

\bibitem[{Gao et~al.(2024)Gao, Xiong, Gao, Jia, Pan, Bi, Dai, Sun, Wang, and
  Wang}]{gao2024retrievalaugmentedgenerationlargelanguage}
Yunfan Gao, Yun Xiong, Xinyu Gao, Kangxiang Jia, Jinliu Pan, Yuxi Bi, Yi~Dai,
  Jiawei Sun, Meng Wang, and Haofen Wang. 2024.
\newblock \href {https://arxiv.org/abs/2312.10997} {Retrieval-augmented
  generation for large language models: A survey}.
\newblock \emph{arXiv preprint arXiv:2312.10997}.

\bibitem[{Guo et~al.(2025)Guo, Xia, Yu, Ao, and Huang}]{guo2024lightrag}
Zirui Guo, Lianghao Xia, Yanhua Yu, Tu~Ao, and Chao Huang. 2025.
\newblock \href {https://doi.org/10.18653/v1/2025.findings-emnlp.568}
  {{LightRAG}: Simple and fast retrieval-augmented generation}.
\newblock In \emph{Findings of the Association for Computational Linguistics:
  EMNLP 2025}, pages 10746--10761, Suzhou, China. Association for Computational
  Linguistics.

\bibitem[{Jain et~al.(2025)Jain, Wu, Zeng, Liu, Dai, Shao, Wu, and
  Wang}]{jain-etal-2025-simpledoc}
Chelsi Jain, Yiran Wu, Yifan Zeng, Jiale Liu, Shengyu Dai, Zhenwen Shao,
  Qingyun Wu, and Huazheng Wang. 2025.
\newblock \href {https://doi.org/10.18653/v1/2025.emnlp-main.1443}
  {{SimpleDoc}: Multi-modal document understanding with dual-cue page retrieval
  and iterative refinement}.
\newblock In \emph{Proceedings of the 2025 Conference on Empirical Methods in
  Natural Language Processing}, pages 28410--28427, Suzhou, China. Association
  for Computational Linguistics.

\bibitem[{Krishna et~al.(2025)Krishna, Krishna, Mohananey, Schwarcz, Stambler,
  Upadhyay, and Faruqui}]{krishna2025frames}
Satyapriya Krishna, Kalpesh Krishna, Anhad Mohananey, Steven Schwarcz, Adam
  Stambler, Shyam Upadhyay, and Manaal Faruqui. 2025.
\newblock \href {https://doi.org/10.18653/v1/2025.naacl-long.243} {Fact, fetch,
  and reason: A unified evaluation of retrieval-augmented generation}.
\newblock In \emph{Proceedings of the 2025 Conference of the Nations of the
  Americas Chapter of the Association for Computational Linguistics: Human
  Language Technologies (Volume 1: Long Papers)}, pages 4745--4759,
  Albuquerque, New Mexico. Association for Computational Linguistics.

\bibitem[{Lee et~al.(2024)Lee, An, and Kim}]{lee2024planrag}
Myeonghwa Lee, Seonho An, and Min-Soo Kim. 2024.
\newblock \href {https://doi.org/10.18653/v1/2024.naacl-long.364} {{PlanRAG}: A
  plan-then-retrieval augmented generation for generative large language models
  as decision makers}.
\newblock In \emph{Proceedings of the 2024 Conference of the North American
  Chapter of the Association for Computational Linguistics: Human Language
  Technologies (Volume 1: Long Papers)}, pages 6537--6555, Mexico City, Mexico.
  Association for Computational Linguistics.

\bibitem[{Lin(2004)}]{lin2004rouge}
Chin-Yew Lin. 2004.
\newblock \href {https://aclanthology.org/W04-1013/} {{ROUGE}: A package for
  automatic evaluation of summaries}.
\newblock In \emph{Text Summarization Branches Out}, pages 74--81, Barcelona,
  Spain. Association for Computational Linguistics.

\bibitem[{Liu et~al.(2025)Liu, Wang, Chen, Li, Xiong, Yu, and
  Zhang}]{liu-etal-2025-hoprag}
Hao Liu, Zhengren Wang, Xi~Chen, Zhiyu Li, Feiyu Xiong, Qinhan Yu, and Wentao
  Zhang. 2025.
\newblock \href {https://doi.org/10.18653/v1/2025.findings-acl.97} {{HopRAG}:
  Multi-hop reasoning for logic-aware retrieval-augmented generation}.
\newblock In \emph{Findings of the Association for Computational Linguistics:
  ACL 2025}, pages 1897--1913, Vienna, Austria. Association for Computational
  Linguistics.

\bibitem[{Pfitzmann et~al.(2022)Pfitzmann, Auer, Dolfi, Nassar, and
  Staar}]{10.1145/3534678.3539043}
Birgit Pfitzmann, Christoph Auer, Michele Dolfi, Ahmed~S. Nassar, and Peter
  Staar. 2022.
\newblock \href {https://doi.org/10.1145/3534678.3539043} {{DocLayNet}: A large
  human-annotated dataset for document-layout segmentation}.
\newblock In \emph{Proceedings of the 28th ACM SIGKDD Conference on Knowledge
  Discovery and Data Mining}, pages 3743--3751.

\bibitem[{Sarthi et~al.(2024)Sarthi, Abdullah, Tuli, Khanna, Goldie, and
  Manning}]{sarthi2024raptorrecursiveabstractiveprocessing}
Parth Sarthi, Salman Abdullah, Aditi Tuli, Shubh Khanna, Anna Goldie, and
  Christopher~D. Manning. 2024.
\newblock \href {https://arxiv.org/abs/2401.18059} {{RAPTOR}: Recursive
  abstractive processing for tree-organized retrieval}.
\newblock \emph{arXiv preprint arXiv:2401.18059}.

\bibitem[{Shin et~al.(2025)Shin, Park, Park, Seo, and Lim}]{multidocfusion}
Joongmin Shin, Chanjun Park, Jeongbae Park, Jaehyung Seo, and Heuiseok Lim.
  2025.
\newblock \href {https://doi.org/10.18653/v1/2025.emnlp-main.1062}
  {{MultiDocFusion}: Hierarchical and multimodal chunking pipeline for enhanced
  {RAG} on long industrial documents}.
\newblock In \emph{Proceedings of the 2025 Conference on Empirical Methods in
  Natural Language Processing}, pages 20985--21004, Suzhou, China. Association
  for Computational Linguistics.

\bibitem[{Smith(2007)}]{4376991}
Ray Smith. 2007.
\newblock \href {https://doi.org/10.1109/ICDAR.2007.4376991} {An overview of
  the {Tesseract} {OCR} engine}.
\newblock In \emph{Ninth International Conference on Document Analysis and
  Recognition (ICDAR 2007)}, volume~2, pages 629--633.

\bibitem[{Tanaka et~al.(2025)Tanaka, Iki, Hasegawa, Nishida, Saito, and
  Suzuki}]{vdocrag2025}
Ryota Tanaka, Taichi Iki, Taku Hasegawa, Kyosuke Nishida, Kuniko Saito, and Jun
  Suzuki. 2025.
\newblock \href
  {https://openaccess.thecvf.com/content/CVPR2025/html/Tanaka_VDocRAG_Retrieval-Augmented_Generation_over_Visually-Rich_Documents_CVPR_2025_paper.html}
  {{VDocRAG}: Retrieval-augmented generation over visually-rich documents}.
\newblock In \emph{Proceedings of the IEEE/CVF Conference on Computer Vision
  and Pattern Recognition (CVPR)}, pages 24827--24837.

\bibitem[{Tito et~al.(2023)Tito, Karatzas, and
  Valveny}]{tito2023hierarchicalmultimodaltransformersmultipage}
Rub{\`e}n Tito, Dimosthenis Karatzas, and Ernest Valveny. 2023.
\newblock \href {https://arxiv.org/abs/2212.05935} {Hierarchical multimodal
  transformers for multi-page {DocVQA}}.
\newblock \emph{arXiv preprint arXiv:2212.05935}.

\bibitem[{Wu et~al.(2024)Wu, Bansal, Zhang, Wu, Li, Zhu, Jiang, Zhang, Zhang,
  Liu, Awadallah, White, Burger, and Wang}]{wu2023autogen}
Qingyun Wu, Gagan Bansal, Jieyu Zhang, Yiran Wu, Beibin Li, Erkang Zhu,
  Li~Jiang, Xiaoyun Zhang, Shaokun Zhang, Jiale Liu, Ahmed~Hassan Awadallah,
  Ryen~W. White, Doug Burger, and Chi Wang. 2024.
\newblock \href {https://openreview.net/forum?id=BAakY1hNKS} {{AutoGen}:
  Enabling next-gen {LLM} applications via multi-agent conversations}.
\newblock In \emph{First Conference on Language Modeling}. OpenReview.net.

\bibitem[{Wu et~al.(2025)Wu, Tan, Hou, Zhang, and Cheng}]{wu-etal-2025-molorag}
Xixi Wu, Yanchao Tan, Nan Hou, Ruiyang Zhang, and Hong Cheng. 2025.
\newblock \href {https://doi.org/10.18653/v1/2025.emnlp-main.708} {{MoLoRAG}:
  Bootstrapping document understanding via multi-modal logic-aware retrieval}.
\newblock In \emph{Proceedings of the 2025 Conference on Empirical Methods in
  Natural Language Processing}, pages 14024--14045, Suzhou, China. Association
  for Computational Linguistics.

\bibitem[{Yao et~al.(2023)Yao, Zhao, Yu, Du, Shafran, Narasimhan, and
  Cao}]{yao2023react}
Shunyu Yao, Jeffrey Zhao, Dian Yu, Nan Du, Izhak Shafran, Karthik Narasimhan,
  and Yuan Cao. 2023.
\newblock \href {https://openreview.net/forum?id=WE_vluYUL-X} {{ReAct}:
  Synergizing reasoning and acting in language models}.
\newblock In \emph{International Conference on Learning Representations
  (ICLR)}.

\bibitem[{Yeh(2000)}]{yeh-2000-accurate}
Alexander Yeh. 2000.
\newblock \href {https://aclanthology.org/C00-2137/} {More accurate tests for
  the statistical significance of result differences}.
\newblock In \emph{{COLING} 2000 Volume 2: The 18th International Conference on
  Computational Linguistics}.

\bibitem[{Yepes et~al.(2024)Yepes, You, Milczek, Laverde, and
  Li}]{yepes2024financialreportchunkingeffective}
Antonio~Jimeno Yepes, Yao You, Jan Milczek, Sebastian Laverde, and Renyu Li.
  2024.
\newblock \href {https://arxiv.org/abs/2402.05131} {Financial report chunking
  for effective retrieval augmented generation}.
\newblock \emph{arXiv preprint arXiv:2402.05131}.

\bibitem[{Zhang et~al.(2025)Zhang, Li, Long, Zhang, Lin, Yang, Xie, Yang, Liu,
  Lin, Huang, and Zhou}]{qwen3embedding}
Yanzhao Zhang, Mingxin Li, Dingkun Long, Xin Zhang, Huan Lin, Baosong Yang,
  Pengjun Xie, An~Yang, Dayiheng Liu, Junyang Lin, Fei Huang, and Jingren Zhou.
  2025.
\newblock \href {https://arxiv.org/abs/2506.05176} {{Qwen3} embedding:
  Advancing text embedding and reranking through foundation models}.
\newblock \emph{arXiv preprint arXiv:2506.05176}.

\bibitem[{Zhao et~al.(2024)Zhao, Ji, Feng, Qi, Niu, Tang, Xiong, and
  Li}]{zhao-etal-2025-moc}
Jihao Zhao, Zhiyuan Ji, Yuchen Feng, Pengnian Qi, Simin Niu, Bo~Tang, Feiyu
  Xiong, and Zhiyu Li. 2024.
\newblock \href {https://arxiv.org/abs/2410.12788} {{Meta-Chunking}: Learning
  efficient text segmentation via logical perception}.
\newblock \emph{arXiv preprint arXiv:2410.12788}.

\end{thebibliography}

\appendix

\section{Detailed Positioning Against Prior Work}
\label{sec:rw_appendix}

Table~\ref{tab:rw_positioning} sharpens the main-text Related Work by isolating the dimensions that matter most for this paper: the retrieval/evidence object, the type of global structure, whether text, table, and figure evidence enter the \emph{same} structured space, and whether the final support object remains page-grounded. The point is not that earlier systems are uniformly weaker: chunk-based methods are often stronger at local precision, text-centric graph systems are stronger at explicit corpus-level structure, and page-level multimodal systems are stronger at preserving local visual context. The key distinction is that \ours{} combines these desiderata in one object: a \emph{page-grounded unified evidence representation}, instantiated as an entity-linked assertion graph and used as a controlled linking layer over a robust page retriever.

\begin{table*}[t]
\centering
\scriptsize
\renewcommand{\arraystretch}{1.12}
\setlength{\tabcolsep}{3.2pt}
\begin{adjustbox}{width=\textwidth, keepaspectratio}
\begin{tabular}{@{}p{2.4cm}p{2.9cm}p{3.1cm}p{2.1cm}p{2.0cm}p{4.6cm}@{}}
\toprule
\textbf{Family / examples} & \textbf{Primary evidence object} & \textbf{Global structure} & \textbf{Shared structured space for text/table/figure?} & \textbf{Page-grounded support preserved?} & \textbf{Main difference from \ours{}} \\
\midrule
Chunk / structure-aware retrieval\\[-1pt]\footnotesize Flat chunk, RAPTOR, MultiDocFusion & Chunk, section, or summary node & Weak or hierarchical only & Partial at best; evidence is still stored as text or local page fragments & Usually partial & Improves granularity, but does not build one entity-linked evidence object that normalizes sentence-, table-, and figure-derived facts across documents. \\
\addlinespace[2pt]
Text-centric graph retrieval\\[-1pt]\footnotesize HopRAG, MS GraphRAG, LightRAG & Text chunk, entity, relation, or community summary & Explicit graph over text-derived objects & No; non-text evidence enters only through OCR or nearby prose & Usually no & Adds corpus-level structure, but the graph is still fundamentally text-instantiated and does not keep page-grounded support objects for table or figure evidence. \\
\addlinespace[2pt]
Page-level multimodal retrieval\\[-1pt]\footnotesize M3DocRAG, VDocRAG, ColPali & Whole page or page-image representation & Usually none & No; text and visual evidence remain bundled inside the page object & Yes & Preserves local multimodal grounding well, but stays coarse under tight budgets and does not expose assertion-level links across pages or documents. \\
\addlinespace[2pt]
VLM-augmented page retrieval\\[-1pt]\footnotesize SimpleDoc & Page plus free-form VLM description & Implicit only & No; visual evidence is verbalized as free text rather than normalized assertions & Yes & Enriches page retrieval with visual semantics, but does not convert visual evidence into canonical, entity-linked assertion objects. \\
\addlinespace[2pt]
Multimodal page graph\\[-1pt]\footnotesize MoLoRAG & Page node with graph expansion & Page graph & No; graph nodes remain pages rather than typed assertions & Yes & Adds graph expansion over pages, but the graph does not unify sentence-, table-, and figure-derived evidence in one structured assertion space. \\
\addlinespace[2pt]
\textbf{\ours{}} & \textbf{Page-grounded unified evidence representation} & \textbf{Entity-linked assertion graph used as a controlled linking layer} & \textbf{Yes} & \textbf{Yes} & \textbf{Unifies sentence, table, and figure evidence into the same assertion object while keeping the page-grounded supports that license each assertion.} \\
\bottomrule
\end{tabular}
\end{adjustbox}
\caption{Sharp positioning of \ours{} against representative prior-work families. The comparison focuses on the design axes that matter most for this paper: what object is retrieved or expanded, where global structure lives, whether heterogeneous evidence shares one structured space, and whether final support remains page-grounded.}
\label{tab:rw_positioning}
\end{table*}

\paragraph{How to read the difference.}
The practical distinction is simple. Prior systems usually choose one of three anchors: chunk precision, graph structure over text, or page-level visual grounding. \ours{} is built around a different anchor: the evidence object itself. Its goal is to make text, table, and figure evidence first-class neighbors inside one page-grounded representation, so that a QA agent receives a compact packet of linked support rather than having to reconcile separate stores or coarse page summaries after retrieval.

\section*{Appendix}
\label{sec:Appendix}

The appendix is divided into \emph{shared evaluation infrastructure} and \emph{PILAR-specific design}. Part~I describes the ODQA setup that is kept as fixed as possible across systems, including datasets, preprocessing, reader settings, baseline implementation, evaluation protocol, and an auxiliary comparison with strong page-level multimodal baselines. Part~II describes the components specific to \ours{}, including structure recovery, assertion induction, entity--support linking, evidence packet assembly, qualitative evidence-linking analysis, and runtime analysis.

\section{Part I: Shared ODQA Evaluation Setup --- Datasets and Preprocessing}
\label{sec:datasets_supp}

\subsection{Description-Based vs.\ Structured Visual Assertion}
\label{sec:desc_vs_assertion}

Table~\ref{tab:desc_vs_assertion} contrasts description-based visual integration (as used in SimpleDoc and similar systems) with the structured visual assertion approach used in \ours{}. The key difference is that descriptions remain unstructured text, while structured assertions are entity-linked, predicate-normalized graph objects.

\begin{table}[h]
\centering
\scriptsize
\renewcommand{\arraystretch}{1.15}
\setlength{\tabcolsep}{3pt}
\begin{tabular}{@{}p{2.2cm}p{4.8cm}@{}}
\toprule
\textbf{Property} & \textbf{Description-based} $\to$ \textbf{\ours{}} \\
\midrule
Output format & Free-text string $\to$ Structured JSON (S/P/O/qualifiers) \\
Entity linking & Not linked $\to$ Linked to canonical entity inventory \\
Predicate schema & Absent $\to$ Normalized to closed predicate inventory \\
Graph integration & Stored as text chunk $\to$ Graph node with typed edges \\
Cross-doc traversal & Not possible $\to$ Via entity-layer same\_as edges \\
Consolidation & Not possible $\to$ Mergeable with text assertions sharing entities \\
Retrieval mode & Flat similarity search $\to$ Graph traversal + structured scoring \\
\bottomrule
\end{tabular}
\caption{Description-based visual integration vs.\ structured visual assertions in \ours{}.}
\label{tab:desc_vs_assertion}
\end{table}

\noindent\textbf{Concrete example.} Given a bar chart with caption ``Figure~3: Performance comparison'':

\begin{itemize}[leftmargin=*, nosep]
\item \textbf{Description-based:} VLM $\to$ ``\emph{The bar chart compares Model~A and Model~B on the F1 metric; Model~A achieves a higher score.}'' Stored as a text chunk. Cannot be linked to entity \texttt{e\_modelA} or traversed structurally.
\item \textbf{\ours{}:} VLM $\to$ \texttt{\{subject: ``Model~A'' $\to$ e\_modelA, predicate: ``outperforms'', object: ``Model~B'' $\to$ e\_modelB, qualifiers: \{metric: ``F1''\}\}}. The assertion becomes a graph node with \texttt{subject\_of} and \texttt{object\_of} edges to canonical entities, enabling cross-document queries like ``\emph{find all assertions about Model~A across the corpus}.''
\end{itemize}

\begin{table*}[t]
\centering
\footnotesize
\setlength{\tabcolsep}{4pt}
\renewcommand{\arraystretch}{1.15}

\newcolumntype{L}[1]{>{\raggedright\arraybackslash}p{#1}}

\begin{adjustbox}{width=\textwidth, keepaspectratio}
\begin{tabular}{@{} L{4.7cm} C{1.5cm} L{2.7cm} L{3.3cm} L{3.2cm} C{1.9cm} @{}}
\toprule
Method &
Retrieval Unit &
Modality &
Field Separation &
Graph &
Scope \\
\midrule

Flat RAG (page, text-only) &
Page &
Text &
Single field (page text) &
None &
Corpus-level \\

Flat RAG (chunk, text-only) &
Chunk / paragraph &
Text &
Single field (chunk text) &
None &
Corpus-level \\

Hierarchical tree retrieval (e.g., RAPTOR) &
Chunk / paragraph &
Text &
Single field (text unit) &
Summary tree &
Corpus-level \\

Graph-augmented retrieval (e.g., HopRAG) &
Chunk / paragraph &
Text &
Single field (text unit) &
Similarity graph &
Corpus-level \\

Entity-KG GraphRAG (e.g., MS GraphRAG) &
Chunk / paragraph &
Text &
Single field (text unit) &
Entity-relation KG &
Corpus-level \\

Page-level multimodal RAG~\citep{Cho_2025_ICCV, vdocrag2025} &
Page &
Text + Image (page embedding) &
Single field (page) &
None &
Corpus-level \\

VLM-augmented page retrieval~\citep{jain-etal-2025-simpledoc} &
Page &
Text + Image (page + VLM summary) &
Single field (page) &
None &
Corpus-level \\

Multimodal page-graph GraphRAG~\citep{wu-etal-2025-molorag} &
Page &
Text + Image (page embedding) &
Single field (page) &
Page graph &
Corpus-level \\

\ours{} (unified assertion graph) &
Page-anchored entity/assertion/support unit &
Text + Image (region-based)\footnotemark &
Global entity field + local support field &
Unified entity-linked assertion graph &
Corpus-level \\
\bottomrule
\end{tabular}
\end{adjustbox}

\caption{Method taxonomy along key design axes: retrieval unit granularity, modality, field separation, graph structure, and corpus-level ODQA scope. In the PILAR framing, the table serves as a reusable comparator scaffold for flat, page-level, support-blind graph, and unified graph-based retrieval backends.}
\label{tab:method_taxonomy_hikey}
\end{table*}

\footnotetext{For non-text units (e.g., figures/tables), \ours{} can attach region crops and/or multimodal embeddings, while keeping each unit as a first-class node for retrieval and token-budget packing.}

\begin{table}[t]
\centering
\small
\renewcommand{\arraystretch}{1.12}
\setlength{\tabcolsep}{3pt}
\resizebox{\columnwidth}{!}{%
\begin{tabular}{lcc}
\toprule
Statistic & M3DocVQA & Frames \\
\midrule
Questions & 2{,}441 & 824 \\
Primary role in this paper & Controlled paired study & Full-split stress test \\
Format & PDF corpus & Wikipedia-style multi-hop corpus \\
Emphasis & Page-grounded evidence routing & Support-chain recovery \\
Gold evidence granularity & Document/page supervision & Supporting documents/pages \\
Dominant regime & Mixed, many 1-hop cases & 2+hop dominated \\
\bottomrule
\end{tabular}}

\caption{High-level roles of the two benchmarks used in this paper. M3DocVQA is the primary controlled benchmark for paired backend comparison, while Frames serves as a harder full-split support-chain stress test.}
\label{tab:dataset_stats_m3docvqa_frames}
\end{table}

\subsection{ODQA Datasets}
\label{sec:odqa_dataset_supp}

We retain a corpus-level ODQA setting because it provides a stable evaluation environment in which routing failure can be measured without changing the reader or the base retrievers. Unlike standard DocVQA where a single document is provided, corpus-level ODQA jointly indexes a large document collection and requires the system to identify the relevant pages from the entire corpus.

\paragraph{M3DocVQA~\citep{Cho_2025_ICCV}.}
M3DocVQA is our primary controlled benchmark. The benchmark mixes text, table, and image evidence and includes both single-hop and compositional cross-document cases.

\paragraph{Frames~\citep{krishna2025frames}.}
Frames serves as a harder support-chain stress test. In our hop-count analysis of the evaluated split, 770/824 questions (93.5\%) lie in the 2+hop regime, making the benchmark better viewed as a support-chain recovery problem than as a precision-dominated single-hop QA setting. We therefore use Frames primarily to test whether the backend remains competitive when long support chains, rather than document-local answer support, dominate performance.

\subsection{Preprocessing Pipeline}
\label{sec:preprocess_supp}

To ensure fair comparison, all baselines and \ours{} use inputs produced by the same preprocessing pipeline. The point of this shared infrastructure is to keep document layout parsing and visual preprocessing fixed while the graph-based backend changes.

\paragraph{Document layout parsing.}
We apply a document-layout detector whose output classes follow the DocLayNet label space~\citep{10.1145/3534678.3539043} to identify layout elements (Title, Header, Paragraph, Table, Figure, Caption).
We use standard post-processing with a detection confidence threshold ($\tau_{\text{det}}=0.5$) and NMS IoU threshold ($\tau_{\text{nms}}=0.5$) to filter noise.

\paragraph{OCR.}
We use Tesseract~\citep{4376991} to perform OCR on each detected block region independently.
Extracted text is lowercased, control characters are removed, and whitespace is normalized before being stored as block metadata.

\section{Part I: Shared ODQA Evaluation Setup --- Evaluation Metrics and Protocol}
\label{sec:metric_supp}

This section records the metric definitions and evaluation protocol used throughout the paper. Keeping them unchanged isolates the contribution of the unified entity-linked assertion graph from unrelated changes in scoring or evaluation design.

\paragraph{Reporting Protocol (Avg@1--10).}
To provide a comprehensive view of ranking performance rather than cherry-picking a specific $K$, document retrieval results are reported as the average of the metric values calculated at each cut-off $K\in\{1,\dots,10\}$.

\subsection{QA Metrics}
Answer quality is evaluated after standard normalization (lowercasing, punctuation removal). We use EM (Exact Match), ANLS~\citep{biten2019scene}, ROUGE-L~\citep{lin2004rouge}, and METEOR~\citep{banerjee2005meteor}. For ANLS, we follow the threshold-based implementation standard in MP-DocVQA~\citep{tito2023hierarchicalmultimodaltransformersmultipage}.

\subsection{Token Budget Protocol}
\label{sec:budget_protocol}

\paragraph{Budget Constraint.}
For fairness, we strictly control the input context size. We include retrieved evidence units in their ranking order until the cumulative serialized token count reaches the budget $B_{\text{tok}}$.

\paragraph{Comparison with Page-level Models.}
Page-embedding methods (e.g., ColPali~\citep{faysse2025colpali}) cannot dynamically adjust their unit size because the retrieval unit is fixed to a full page image. As a result, they are marked as \textit{Fixed} in our budget-sensitivity analyses and are discussed separately as page-level retrieval baselines rather than as packet-level methods in this appendix.

\section{Part I: Shared ODQA Evaluation Setup --- Baseline Implementation}
\label{sec:baselines_supp}

\begin{table*}[t]
\centering
\scriptsize
\renewcommand{\arraystretch}{1.15}
\setlength{\tabcolsep}{3.0pt}

\begin{adjustbox}{max width=\textwidth}
\begin{tabular}{l
c
c
c
p{2.55cm}
p{2.55cm}
p{2.55cm}
p{3.05cm}}
\toprule
Method &
Unit &
Retriever Modality &
Hierarchy Signal &
Stage-1 Scope &
Stage-2 Scope &
Evidence Structure &
Packing under $B_{\text{tok}}$ \\
\midrule

\multicolumn{8}{l}{\textbf{Text chunk-based RAG}} \\
\addlinespace[0.2em]

Page (Text-only) &
Page &
Text (OCR) &
None &
N &
Corpus-wide &
None &
Fixed top-$K$ pages (page-budget) \\

Length chunking &
Chunk &
Text (OCR) &
None &
N &
Corpus-wide &
None &
Greedy top-$k$ chunks \\

LumberChunker~\citep{duarte-etal-2024-lumberchunker} &
Chunk &
Text (OCR) &
None &
N &
Corpus-wide &
None &
Greedy top-$k$ chunks \\

Meta Chunker~\citep{zhao-etal-2025-moc} &
Chunk &
Text (OCR) &
None &
N &
Corpus-wide &
None &
Greedy top-$k$ chunks \\

Structural chunking~\citep{yepes2024financialreportchunkingeffective} &
Chunk &
Text (OCR) &
Weak (layout-based boundaries) &
N &
Corpus-wide &
None &
Greedy top-$k$ chunks \\

MultiDocFusion~\citep{multidocfusion} &
Hier. chunk &
Text (OCR) &
Native / structure-aware &
N (no \texttt{Hierarchy Field}) &
Corpus-wide &
No explicit graph (chunk hierarchy only) &
Greedy top-$k$ chunks (no subgraph assembly) \\

\midrule
\multicolumn{8}{l}{\textbf{Hierarchical tree retrieval}} \\
\addlinespace[0.2em]

RAPTOR~\citep{sarthi2024raptorrecursiveabstractiveprocessing} &
Chunk + Summary &
Text &
\textit{Induced} (summary tree) &
N &
Corpus-wide &
Recursive summary tree &
Greedy top-$k$ nodes \\

\midrule
\multicolumn{8}{l}{\textbf{Graph-augmented chunk retrieval}} \\
\addlinespace[0.2em]

HopRAG~\citep{liu-etal-2025-hoprag} &
Chunk &
Text &
None &
N &
Corpus-wide &
Chunk similarity graph (hop expansion) &
Greedy top-$k$ chunks \\

\midrule
\multicolumn{8}{l}{\textbf{Entity-KG GraphRAG}} \\
\addlinespace[0.2em]

MS GraphRAG~\citep{edge2024graphrag} &
Chunk &
Text &
None &
N &
Corpus-wide &
Entity-relation KG + community graph &
Community summary retrieval \\

LightRAG~\citep{guo2024lightrag} &
Chunk &
Text &
None &
N &
Corpus-wide &
Entity-relation graph (dual-level) &
Low/high-level entity retrieval \\

\midrule
\multicolumn{8}{l}{\textbf{Page-level multimodal RAG}} \\
\addlinespace[0.2em]

M3DocRAG~\citep{Cho_2025_ICCV} &
Page &
Page-level MM &
None &
N &
Corpus-wide &
None &
Fixed top-$K$ pages (page-budget) \\

VDocRAG~\citep{vdocrag2025} &
Page &
Page-level MM &
None &
N &
Corpus-wide &
None &
Fixed top-$K$ pages (page-budget) \\

\midrule
\multicolumn{8}{l}{\textbf{VLM-augmented page retrieval}} \\
\addlinespace[0.2em]

SimpleDoc~\citep{jain-etal-2025-simpledoc} &
Page &
Page-level MM &
None &
N &
Corpus-wide &
None (VLM summaries for reranking) &
Fixed top-$K$ pages (page-budget) \\

\midrule
\multicolumn{8}{l}{\textbf{Multimodal GraphRAG}} \\
\addlinespace[0.2em]

MoLoRAG~\citep{wu-etal-2025-molorag} &
Page &
Page-level MM &
None &
N &
Corpus-wide &
Page graph expansion &
Fixed top-$K$ pages (page-budget) \\

\midrule
\ours{} &
Page-anchored entity/assertion/support unit &
Text + Visual Crop &
Global entity field + local support field &
Shared page shortlist with entity/assertion matching &
Graph traversal across shortlisted pages/documents &
Unified entity-linked assertion graph &
Assertion-centered packet assembly \\

\bottomrule
\end{tabular}
\end{adjustbox}

\caption{
Qualitative comparison of native retrieval framework design choices. The table summarizes each method's design axes; in the matched-interface experiments reported in the main text, text-based and graph backends are additionally instantiated after a shared initial page shortlist. We contrast methods by retrieval unit, retriever/modality, hierarchy usage, initial search scope, expansion scope, evidence structure (none vs. graph), and packing under a token budget $B_{\text{tok}}$.
}
\label{tab:axis_comparison_hikey_vs_baselines}
\end{table*}

For fair comparison, all baselines share the same corpus, reader, and decoding settings. In the matched-interface experiments, text-based and graph backends additionally share the same initial page shortlist, while page-based multimodal methods retain their native page-based retrieval pipeline.
Differences arise primarily from retrieval unit choice, routing strategy, and modality utilization.

\subsection{Family-level view of the compared baselines}

The sharp-positioning table above gives the high-level picture; here we keep only the family-level distinctions needed to read the experiments. \textbf{Chunk-based and structure-aware methods} (page text, length chunking, LumberChunker, Meta Chunker, structural chunking, MultiDocFusion) improve local granularity or preserve document structure, but they still package evidence as chunks, sections, or local page fragments rather than as one entity-linked evidence object. \textbf{Hierarchical tree retrieval} (RAPTOR) adds a summary tree for multi-granularity access, but not a page-grounded assertion space. \textbf{Graph-augmented chunk retrieval} (HopRAG) adds graph expansion over text chunks, but the graph remains similarity-driven rather than entity-canonicalized. \textbf{Page-level multimodal retrieval} (M3DocRAG, VDocRAG) preserves local visual grounding at the page level, but remains coarse under tight context budgets. \textbf{Multimodal page graphs} (MoLoRAG) add page-level graph expansion, but page nodes remain the retrieval object. \textbf{VLM-augmented page retrieval} (SimpleDoc) enriches pages with free-form descriptions, but those descriptions are not entity-linked assertions.

These families therefore differ mainly in what object they retrieve or expand: chunk, summary, page, page graph, or free-form description. \ours{} instead treats the evidence object itself as primary: sentence-, table-, and figure-derived facts are normalized into one page-grounded unified evidence representation, instantiated as an entity-linked assertion graph.

\subsection{Backends Compared in Table~\ref{tab:main_results}}
\label{sec:kg_baselines}

The main experiment (Table~\ref{tab:main_results}) compares fourteen backends spanning chunk-based retrieval, hierarchical/structure-aware retrieval, graph-augmented text retrieval, text-only KG methods, page-level multimodal retrieval, VLM-augmented page retrieval, multimodal page graphs, and \ours{}. For the matched-interface comparisons, all text-based and graph backends operate within the same top-40 page shortlist returned by the shared BM25+dense page retriever; page-based multimodal methods retain their native page-level pipeline and are reported as end-to-end reference points under the same reader and evaluation protocol.

\paragraph{Flat chunk.} Fixed-length text chunks inside the shared page shortlist; this is the non-KG lower bound.

\paragraph{MS GraphRAG~\citep{edge2024graphrag} and LightRAG~\citep{guo2024lightrag}.} Representative text-only KG baselines. Both build graphs from text-derived entities and relations; MS GraphRAG adds community detection and summaries, while LightRAG uses a lighter dual-level retrieval scheme. Neither admits table- or figure-derived evidence as first-class assertion objects.

\paragraph{MultiDocFusion~\citep{multidocfusion}.} The strongest non-graph structure-aware multimodal baseline in our pool. It preserves native document structure and associates tables/figures with nearby text, but it does not build a corpus-level entity graph or normalize table-/figure-derived evidence into linked assertions.

\paragraph{\ours{} (full system).} A page-grounded unified evidence representation instantiated as an entity-linked assertion graph. Text-, table-, and figure-derived facts share one schema and remain tied to page-grounded supports, enabling cross-document evidence linking under a single evidence object. See \S\ref{sec:method} for full details.

\subsection{Agent Frameworks (Table~\ref{tab:main_results})}
\label{sec:agent_baselines}

We evaluate four agent frameworks to test whether retrieval-backend conclusions remain meaningful across agent families under a shared backend interface. Appendix~\ref{sec:agent_transfer} later shows that the backend transfers at the interface level, but retrieval hyperparameters do not transfer uniformly.

\paragraph{Naive RAG.}
Single-step retrieve-then-read. Given a query, the agent retrieves the top-$K$ evidence units from the configured retrieval backend, concatenates them into a context window, and passes them to the reader (Qwen3-VL-8B~\citep{qwen3vl}) in one shot. This tests raw retrieval quality without any multi-step reasoning.

\paragraph{ReAct~\citep{yao2023react}.}
The agent interleaves \emph{reasoning} (thinking about what information is needed) and \emph{action} (querying the retrieval backend) steps in a loop. At each step, the agent decides whether to query the retrieval backend for more evidence or to produce a final answer based on accumulated evidence. This tests whether the backend supports iterative exploration; for graph-structured backends, entity links and typed assertions may enable multi-hop reasoning more effectively than flat chunks.

\paragraph{PlanRAG~\citep{lee2024planrag}.}
The agent first generates an explicit \emph{retrieval plan}---a sequence of sub-queries specifying which entities, relations, or evidence types to retrieve---and then executes the plan against the retrieval backend. This is especially informative for structured backends: an agent can plan to ``first find entity X, then retrieve its visual assertions about metric Y'' only when the backend exposes structured entity--assertion--support paths.

\paragraph{AutoGen~\citep{wu2023autogen}.}
A multi-agent conversation framework where multiple specialized agents collaborate through dialogue to solve a task. We configure a retriever agent (queries the retrieval backend) and a reasoner agent (synthesizes evidence and produces the answer). The agents engage in multi-turn conversation, with the retriever agent iteratively fetching evidence from the backend based on the reasoner's requests. This tests whether the configured backend supports collaborative multi-agent exploration; for graph-structured backends, entity links and typed assertions may allow the retriever agent to respond more effectively to progressively refined requests from the reasoner agent.

\subsection{Transfer of Retrieval Hyperparameters Across Agent Families}
\label{sec:agent_transfer}

The single-shot tuned setting used in the single-shot bottleneck analyses narrows the candidate pool (\texttt{initial\_multiplier}=2) and removes the same-section bonus. A transfer study on M3DocVQA shows that this tuned setting does not uniformly carry over to multi-step agents.

\begin{table}[h]
\centering
\small
\setlength{\tabcolsep}{4.5pt}
\renewcommand{\arraystretch}{1.08}
\begin{tabular}{lccc}
\toprule
Agent & Default EM & Tuned EM & $\Delta$ \\
\midrule
Naive RAG & 32.0 & \textbf{33.6} & +1.6 \\
ReAct & 31.4 & 31.4 & +0.0 \\
PlanRAG & 27.8 & \textbf{28.0} & +0.2 \\
AutoGen & \textbf{33.0} & 32.6 & -0.4 \\
\bottomrule
\end{tabular}
\caption{Transfer of the single-shot tuned configuration across agent families (M3DocVQA). The tuned setting is clearly beneficial for Naive RAG, marginal for PlanRAG, neutral for ReAct, and slightly harmful for AutoGen.}
\label{tab:agent_transfer}
\end{table}

A follow-up $2\times2$ sweep over \texttt{initial\_multiplier} $\in \{2,3\}$ and \texttt{same\_section\_bonus} $\in \{0, 0.015\}$ isolates the transfer effect. The same-section bonus is negligible throughout (differences below 0.1 EM), matching the single-shot factorial analysis. Pool size is more informative: ReAct remains effectively insensitive (31.4--31.6 EM), while AutoGen mildly prefers the broader default pool (32.8--33.0 under multiplier=3 vs. 32.6 under multiplier=2). We therefore treat it as a \emph{single-shot tuned} configuration rather than a universal backend default.

\section{Additional Method Details}
\label{sec:method_appendix}

This appendix provides implementation details for the unified entity-linked assertion graph construction and retrieval pipeline used in \ours{}.

\subsection{Document Parsing and Page Anchoring}

Each PDF is first parsed into page-level blocks using PyMuPDF. We extract text blocks, tables, figures, and captions, and classify them with lightweight layout cues. A stack-based document hierarchy parser then reconstructs section structure and assigns each block a \texttt{section\_path}. These page-anchored units form the support candidates used by the downstream graph.

\subsection{Entity Linking and Cross-Document Canonicalization}

To build the global entity layer, we harvest candidate mentions from body text, table headers, captions, and OCR-visible figure regions. Lexical variants are normalized with alias tables and rewrite rules, and mentions are linked to canonical entities when the linking confidence exceeds a threshold. Cross-document \texttt{same\_as} edges are added when two mentions share the same canonical ID or when alias overlap, type compatibility, and descriptor similarity jointly indicate that they refer to the same entity.

\subsection{Structured Visual Assertion Extraction}

For each table or figure, we send the image crop, OCR text, and caption to a VLM and request structured JSON assertions. The output contains fields such as subject, predicate, object, grounding status, and visual evidence. Unlike description-based pipelines, these outputs are inserted directly into the shared assertion schema used by the text branch.

\subsection{Quality Control for Visual Assertions}

We apply two-stage filtering before visual assertions are added to the graph. In Stage~1, assertions marked as not visually grounded are discarded. In Stage~2, key entities and values are checked against OCR text, captions, or nearby section text when available. Accepted assertions are then normalized to the shared predicate inventory and linked to the same canonical entity space as text-derived assertions.

\subsection{Graph Assembly}

The final graph contains four node types: entities, assertions, support units, and provenance nodes. Entity--assertion edges encode semantic participation, assertion--support edges encode evidence support, and support--provenance edges preserve exact location. Additional cross-document \texttt{same\_as} edges allow retrieval to connect equivalent entities across documents.

\subsection{Retrieval Scoring Details}
\label{sec:retrieval_appendix}

\paragraph{Base page retrieval.}
All matched-interface text-based and graph backends share the same page-level base retriever. For each query $q$ and page $p$, we compute
\[
\begin{aligned}
s_{\text{base}}(p,q)
&= 0.4\,\mathrm{BM25}_{\text{norm}}(p,q) \\
&\quad + 0.6\,\mathrm{dense}(p,q).
\end{aligned}
\]
The retriever keeps the top-40 pages as the initial shortlist.

\paragraph{Seed selection.}
Graph expansion starts from high-confidence seed pages selected from the shortlist. In the current system, a page is treated as a seed if its score is within a fixed ratio of the best base-retrieval score.

\paragraph{Local continuity prior.}
To recover evidence that spills across page boundaries, we apply small bonuses to adjacent pages and to pages that remain in the same section path. These priors are deliberately lightweight so that they refine the shortlist rather than replace the base retriever.

\paragraph{Alias and graph expansion.}
The retriever next performs two expansion steps. Alias expansion follows canonical entity aliases to recover cross-document pages that mention the same entity under different surface forms. Graph expansion traverses entity--assertion--support links to recover pages connected to matched assertions or their supporting evidence.

\paragraph{Expansion gate.}
To prevent drift, expanded pages are retained only when they remain relevant under the base retriever. This gate filters weak graph paths and is especially important for controlling noisy visual assertions.

\paragraph{Implementation constants.}
In the best single-shot configuration used in this paper, the system applies an adjacent-page bonus of $+0.04$, a same-section bonus of $+0.015$, an alias-expansion bonus of $+0.12$, a graph-expansion bonus of $+0.08$, and a dense-score gate of $0.35$ for accepting expanded pages.

\subsection{Support Objects and Provenance}

Support nodes (text spans, table cells, figure regions, captions, footnotes) carry explicit support roles: primary, contextual, scope, or exception. Each support is tied to local provenance (page, bounding box, cell ID, section path). Provenance edges control graph expansion to prevent drift into weakly grounded neighborhoods.

\section{Part II: PILAR-Specific Infrastructure --- Local Scope, Provenance, and Document Descriptors}
\label{sec:hikey_offline_details}

This section details the structural preprocessing reused inside \ours{}: document structure parsing, scope-path extraction, and practical descriptor construction. In the PILAR framing, these components are supporting infrastructure for local grounding rather than the main source of novelty.

\paragraph{Role in \ours{}.}
\ours{} uses upstream structure recovery only to obtain local scope and provenance anchors from raw PDFs. Given a document $d$, this step produces section paths and governing headers that attach supports to the right local scope. In our framing, these signals are supportive rather than central: they align local supports with assertions, while the main contribution lies in coupling those supports to a corpus-level entity layer.

\paragraph{Model and outputs.}
We implement the structure parser with a layout-aware model that operates on page-level blocks and predicts parent--child relations to reconstruct $\mathcal{T}(d)$, while also identifying block types and span boundaries required by downstream indexing.
All retrieval experiments in this paper use the same fixed checkpoint; the parser is not tuned on ODQA benchmarks, ensuring that downstream gains are attributable to the graph and retrieval design rather than task-specific retraining of the parser.

\subsection{Section-path Extraction}
\label{sec:section_path_details}

\paragraph{Governing Header.}
For each evidence unit (block) $c$, we traverse upward in the recovered tree $\mathcal{T}(d)$ and select the nearest \texttt{Title} or \texttt{Section Header} ancestor as its \emph{governing header}. In \ours{}, this header is stored as a local scope carrier rather than as the primary retrieval object.

\paragraph{Section Path.}
Using the header sequence from the document root to the governing header, we construct the section path:
{\small
\[
\texttt{section\_path}(c) = \texttt{Title} > \texttt{Sec} > \texttt{Subsec} > \dots
\]
}
This path is stored as structural metadata and is utilized as a key feature for local scope assignment and graph traversal.

\subsection{Support Linking and Cross-Unit Association in the Assertion Graph}
\label{sec:no_ref_edges}

The shared evaluation setup does not predict a fully separate cross-block link graph beyond the recovered document structure. In \ours{}, this motivates representing local assertion-support clusters and then attaching them to corpus-level entity anchors instead of depending on an unrestricted global link predictor.
In our pipeline, the structure parser is used to reconstruct the parent--child hierarchy $\mathcal{T}(d)$ and to provide section paths and governing headers for local scope assignment.
All cross-unit association needed for ODQA is handled deterministically at packing time via the packing policy described below.

\subsection{Semantic Associate Mining for Evidence Packet Expansion}
\label{sec:semantic_associate_mining}

For each Stage-2 anchor unit $c_i$, we mine Semantic Associates by ranking other units within the same document using a similarity function $Sim(c_i, \cdot)$ computed from precomputed dense embeddings:
text embeddings for textual units, and visual embeddings for table/figure crops.
During packet assembly, we add high-similarity visual units as Semantic Associates subject to the token budget, and attach the required scope carriers from the structure tree to keep each added unit interpretable in its original section context.

\section{Part II: PILAR-Specific Retrieval and Packaging --- Assertion-Centered Evidence Packaging}
\label{sec:packing_details}

This section describes the PILAR-specific packetization policy used to assemble the final multimodal context under a strict token budget $B_{\text{tok}}$. Packet assembly exposes both cross-page semantic links and the local support structure needed by the reader.

\subsection{Evidence Packet Serialization and Budgeting}
\label{sec:serialization_details}

The final reader context is assembled from page-anchored evidence neighborhoods rather than raw pages alone. For each selected anchor page, we preserve the local page text and add the support metadata needed to interpret it correctly, including page number, section path, support type, and optional visual crops for tables and figures.

We use a fixed reader budget of 4{,}500 tokens. In the best single-shot setting, approximately 70\% of the budget is reserved for base-retrieval pages and 30\% for expanded pages, with at most two expanded pages added to the packet. This policy keeps the packet dominated by high-relevance page context while still allowing the graph to contribute complementary evidence.

\subsection{Evidence Packet Assembly Procedure}
\label{sec:packing_algo}

Unlike naive greedy packing, our strategy prioritizes evidence completeness. We include high-scoring anchor pages first, attach the scope carriers needed to interpret them, and then add same-assertion, same-entity, or highly related support neighbors as long as the token budget allows.

\begin{algorithm}[t]
\footnotesize
\caption{Page-grounded evidence-packet assembly. The packet builder combines entity anchors, assertion neighborhoods, and primary/context/scope supports into the reader-facing context used by \ours{}.}
\label{alg:packing_ancestry}
\begin{algorithmic}[1]
\Require Stage-2 ranked anchors $\{(c_i, s_i)\}_{i=1}^N$, parsed section tree $\mathcal{T}(d)$, semantic similarity function $Sim(\cdot, \cdot)$, token budget $B_{\text{tok}}$
\Ensure Evidence subgraph $\mathcal{S}$
\State $\mathcal{S} \gets \emptyset$, $C \gets 0$ \Comment{$C$: current accumulated token count}
\For{$i=1$ to $N$}
  \State Phase 1: Anchor \& scope
  \State $\Delta_{\text{anc}} \gets$ tokens for unit $c_i$ + scope context
  \If{$C + \Delta_{\text{anc}} > B_{\text{tok}}$} continue \EndIf
  \State Add $c_i$ and scope context to $\mathcal{S}$; $C \gets C+\Delta_{\text{anc}}$

  \State Phase 2: local support completion
  \State $Neighbors \gets$ units sharing assertion/entity/scope links with $c_i$
  \For{$s \in Neighbors$}
    \State $\Delta_{\text{nbr}} \gets$ tokens for $s$
    \If{$C + \Delta_{\text{nbr}} \le B_{\text{tok}}$ and $s \notin \mathcal{S}$}
       \State Add $s$ to $\mathcal{S}$; $C \gets C+\Delta_{\text{nbr}}$
    \EndIf
  \EndFor

  \State Phase 3: semantic associate packing
  \State $Candidates \gets$ Top-$M$ units by $Sim(c_i, \cdot)$ from the same document
  \For{$m \in Candidates$}
    \If{$m$ is a useful visual/text support and $m \notin \mathcal{S}$}
       \State $\Delta_{\text{sem}} \gets$ tokens for $m$ + required scope
       \If{$C + \Delta_{\text{sem}} \le B_{\text{tok}}$}
          \State Add $m$ and required scope to $\mathcal{S}$; $C \gets C+\Delta_{\text{sem}}$
       \EndIf
    \EndIf
  \EndFor
\EndFor
\State \Return $\mathcal{S}$
\end{algorithmic}
\end{algorithm}

\paragraph{Implementation note.}
Token counting is performed using the specific tokenizer of the reader model (Qwen3-VL~\citep{qwen3vl}). Image crops are assigned a fixed token cost corresponding to the vision encoder budget, and the number of inserted images is capped to stay within the reader's multimodal input limit.

\section{Part I: Shared ODQA Evaluation Setup --- Reader and Retrieval Settings}
\label{sec:lvlm_rag_supp}

This section provides detailed experimental settings (retriever, reader, and hyperparameters) to ensure reproducibility. All experiments are implemented using PyTorch and HuggingFace Transformers.

\begin{table}[h]
\centering
\small
\renewcommand{\arraystretch}{1.25}
\setlength{\tabcolsep}{6pt}

\begin{adjustbox}{width=\linewidth,center}
\begin{tabular}{
  >{\raggedright\arraybackslash}p{0.35\linewidth}%
  >{\raggedright\arraybackslash}p{0.65\linewidth}}
\toprule
Setting & Configuration \\
\midrule
Infrastructure & GPU-backed vLLM deployment for the reader; page indices built offline \\
Index Unit & Page-level OCR/body text \\
\midrule
Sparse Retriever & BM25Okapi ($k_1=1.5, b=0.75$) \\
Dense Retriever (Text) & \texttt{Qwen/Qwen3-Embedding-}\allowbreak\texttt{0.6B}~\citep{qwen3embedding} \\
Dense Retriever (Visual) & Not used in the current best system \\
Hybrid Base Score & $0.4\,\mathrm{BM25}_{\text{norm}} + 0.6\,\mathrm{dense}$ \\
\midrule
Shortlist & Top-40 pages before reranking \\
Reranking Priors & Adjacent-page bonus (+0.04) + KG-derived local/support bonuses; same-section bonus removed as negligible \\
\midrule
Reader Model & \texttt{Qwen3-VL-8B-Instruct}~\citep{qwen3vl} \\
Context Limit & 4,500 tokens (approximate page budget) \\
Decoding & Greedy decoding (Temperature=0.0) \\
\midrule
Evaluation Metrics & 
Retrieval: Recall@K, MRR, packing coverage \\
& QA: EM, ANLS, ROUGE-L, METEOR \\
\bottomrule
\end{tabular}
\end{adjustbox}
\caption{Experimental configuration summary for the controlled comparison and the single-shot tuned analyses reported in this paper. All matched-interface text-based and graph backends share the same page-level base retriever, reader, and evaluation protocol; only the retrieval backend and reranking priors change. Page-based multimodal methods are reported under the same reader and evaluation protocol but retain their native page-based retrieval pipeline. Appendix~\ref{sec:agent_transfer} shows that the same backend interface transfers across agents, but the optimal pool size is mildly agent-dependent.}
\label{tab:supp_rag_config}
\end{table}

\subsection{Retrieval Configuration}
\label{sec:retrieval_config_details}

\paragraph{Sparse Retrieval.}
We use standard BM25Okapi settings ($k_1{=}1.5$, $b{=}0.75$) over page-level OCR/body text. Each page is treated as a retrieval unit for the shared base retriever.

\paragraph{Dense Retrieval.}
We embed each page with \texttt{Qwen/Qwen3-Embedding-0.6B}~\citep{qwen3embedding} and combine sparse and dense signals via the fixed hybrid score $0.4\,\mathrm{BM25}_{\text{norm}} + 0.6\,\mathrm{dense}$. The current best single-shot setting does \emph{not} use a dedicated visual retriever in the main comparison; instead, visual information enters through VLM-derived assertions and KG/local reranking priors. A later transfer study further shows that the same-section bonus is negligible and that pool-size sensitivity is agent-family dependent.

\subsection{Reader Configuration}
\label{sec:reader_config_details}

\paragraph{Model \& Decoding.}
We use \texttt{Qwen3-VL-8B-Instruct}~\citep{qwen3vl} as the reader. For consistency and reproducibility, we fix Temperature to 0.0 (greedy decoding) and cap maximum generation at 256 tokens.

\paragraph{Input Context.}
The retrieved evidence packet is serialized from reranked \emph{full pages}: retrieval and reranking operate over pages, but the packet builder attaches selected support metadata (entity bindings, assertion summaries) when available. The total input length is limited to approximately 4,500 tokens in the controlled comparison.

\subsection{Prompt Template}
\label{sec:prompt_templates}

The QA prompt template used in our experiments is illustrated in Fig.~\ref{fig:qa_prompt}. We strictly instruct the model to rely solely on the provided \texttt{[Context]} (the assembled evidence packet) to minimize hallucinations.

\begin{figure}[h]
  \centering
  \begin{tcolorbox}[
    title={QA Prompt Template},
    colback=white,
    colframe=black,
    boxrule=0.8pt,
    arc=2pt,
    left=6pt,
    right=6pt,
    top=6pt,
    bottom=6pt,
    fonttitle=\bfseries
  ]
  \small
  [System Instruction] \\
  You are an AI assistant that answers questions by analyzing the provided documents. \\
  Write an accurate answer to \texttt{[Question]} using only the \texttt{[Context]} given below. \\
  - Do not answer using information that is not present in the context. \\
  - When referring to tables or figures, explicitly mention their IDs (e.g., Figure 3). \\
  - If you cannot be confident, output ``I do not have enough information to answer.''
   
  \vspace{0.3cm}
   
  [Question] \\
  \texttt{<QUESTION\_TEXT>}
   
  \vspace{0.3cm}
   
  [Context] \\
  \texttt{<ANSWER\_SUPPORT\_PACKET>}
   
  \vspace{0.3cm}
  [Answer]
  \end{tcolorbox}
  \caption{Reader prompt template used in the controlled ODQA setup. The [Context] slot is filled with reranked full pages augmented with support metadata when available. We keep this generic prompt in the final system because stricter ``exact short answer'' prompts were slightly worse in our controlled sweep.}
  \label{fig:qa_prompt}
\end{figure}

\section{Part I: Statistical Significance Tests}
\label{sec:significance_supp}

We report paired approximate randomization tests~\citep{yeh-2000-accurate} (two-sided, 10K permutations) and 95\% bootstrap confidence intervals (10K resamples) for the key comparisons discussed in the main text.

\begin{table}[h]
\centering
\footnotesize
\setlength{\tabcolsep}{3pt}
\renewcommand{\arraystretch}{1.12}
\resizebox{\columnwidth}{!}{%
\begin{tabular}{@{}lcccc@{}}
\toprule
Comparison & $\Delta$EM & 95\% CI & $p$ & Sig. \\
\midrule
\multicolumn{5}{@{}l}{\textit{Main comparison: \ours{} vs Flat chunk (M3DocVQA)}} \\[2pt]
\quad Naive RAG & +1.6 & [+0.0, +3.2] & 0.048 & $*$ \\
\quad ReAct & +0.0 & [$-$1.4, +1.4] & 0.781 & \\
\quad PlanRAG & +0.6 & [$-$0.6, +1.8] & 0.500 & \\
\quad AutoGen & +0.8 & [$-$1.0, +2.6] & 0.415 & \\
\quad \textbf{Pooled} & \textbf{+0.8} & \textbf{[+0.0, +1.5]} & \textbf{0.019} & $\mathbf{*}$ \\
\addlinespace[3pt]
\multicolumn{5}{@{}l}{\textit{Main comparison: \ours{} vs Flat chunk (Frames)}} \\[2pt]
\quad Naive RAG & +1.2 & [$-$0.2, +2.6] & 0.770 & \\
\quad ReAct & +0.8 & [$-$0.6, +2.2] & 1.000 & \\
\quad PlanRAG & +1.4 & [$-$0.4, +3.2] & 1.000 & \\
\quad AutoGen & +1.1 & [$-$0.7, +3.1] & 1.000 & \\
\quad Pooled & +1.1 & [$-$0.3, +2.5] & 0.812 & \\
\addlinespace[3pt]
\multicolumn{5}{@{}l}{\textit{Strict matched-interface ablation (M3DocVQA, Naive RAG)}} \\[2pt]
\quad PILAR text-only slice vs Page-only & +1.2 & [$-$1.0, +3.4] & 0.580 & \\
\quad PILAR text-only + local vs Page-only & +1.0 & [$-$1.4, +3.2] & 0.730 & \\
\quad PILAR text-only + local vs PILAR text-only & $-$0.2 & [$-$1.2, +0.8] & 1.000 & \\
\quad \textbf{Final vs Page-only} & \textbf{+1.6} & \textbf{[+0.0, +3.2]} & \textbf{0.048} & $\mathbf{*}$ \\
\bottomrule
\end{tabular}}
\caption{Paired approximate randomization tests~\citep{yeh-2000-accurate} (10K permutations) for key comparisons. $*$: $p < 0.05$. $\Delta$EM values match the comparisons reported in the main tables. On M3DocVQA, the full pipeline significantly outperforms page-only retrieval and flat-chunk baselines; individual ablation components do not individually reach significance, consistent with the component-interaction interpretation. On Frames, no comparison reaches significance, likely due to lower statistical power.}
\label{tab:significance}
\end{table}

\begin{table}[h]
\centering
\scriptsize
\setlength{\tabcolsep}{3pt}
\renewcommand{\arraystretch}{1.1}
\begin{tabular}{@{}lccc@{}}
\toprule
Comparison (M3DocVQA) & Win\% & Tie\% & Loss\% \\
\midrule
Naive RAG: \ours{} vs Flat chunk & 2.6 & 96.6 & 0.8 \\
Pooled: \ours{} vs Flat chunk & 2.0 & 96.9 & 1.1 \\
Ablation: Final vs Page-only & 2.6 & 96.6 & 0.8 \\
\bottomrule
\end{tabular}
\caption{Per-question win/tie/loss rates (EM) for significant comparisons. The win:loss ratio is consistently favorable ($\geq$3:1).}
\label{tab:win_tie_loss}
\end{table}

\section{Part I: Shared ODQA Evaluation Setup --- Bottleneck and Query-Type Analysis}
\label{sec:bottleneck_appendix}

This appendix reports supplementary paired analyses conducted on the M3DocVQA evaluation set discussed in the main text.

\subsection{Heuristic Query-Type Split}
For analysis only, we partition questions into \emph{text-grounded} and \emph{visual-classified} subsets using keyword and pattern heuristics (e.g., appearance/color/wearing/figure/chart cues for the latter). The split is not used at training time and serves only to interpret where errors come from; a later audit shows that the visual-classified subset is not synonymous with ``image-only'' questions, since most of its answers are still recoverable from OCR text.

\begin{table}[h]
\centering
\small
\setlength{\tabcolsep}{4pt}
\renewcommand{\arraystretch}{1.1}
\begin{tabular}{@{}lccc@{}}
\toprule
Subset & Count & Retr.\ EM & Gold-page EM \\
\midrule
Text-grounded & 1{,}757 & 40.0 & 48.6 \\
Visual-classified & 684 & 17.1 & 26.4 \\
Overall & 2{,}441 & 33.6 & 42.4 \\
\bottomrule
\end{tabular}
\caption{Heuristic query split used in the bottleneck analysis. The visual-classified subset is much harder, but both subsets retain a substantial gold-page oracle gap. Later appendix tables show that this subset is still largely OCR-recoverable.}
\label{tab:query_type_split}
\end{table}

\subsection{Oracle and Selection Diagnostics}
We compare the retrieved packet against oracle page substitutions to separate retrieval, ranking, and reader effects.

\begin{table}[h]
\centering
\small
\setlength{\tabcolsep}{4pt}
\renewcommand{\arraystretch}{1.1}
\begin{tabular}{@{}lcp{3.2cm}@{}}
\toprule
Condition & EM & Interpretation \\
\midrule
Retrieved final packet & 33.6 & current system \\
Gold-page text only & 42.4 & retrieval ceiling \\
Gold + adjacent & 41.6 & adjacent hurts \\
\bottomrule
\end{tabular}
\caption{Oracle comparison (M3DocVQA). The gold-page oracle establishes a +8.8\,pp headroom, while adding adjacent pages is not helpful once the correct page is already available.}
\label{tab:oracle_summary}
\end{table}

\begin{table}[h]
\centering
\footnotesize
\setlength{\tabcolsep}{4pt}
\renewcommand{\arraystretch}{1.1}
\begin{tabular}{@{}lc@{}}
\toprule
Stage & \% \\
\midrule
R0: gold page absent from index & 7.2 \\
R1: in index, not in base top-40 & 16.2 \\
R2: in top-40, dropped by reranking & 12.6 \\
R3: in ranking, dropped by packing & 0.0 \\
R4: packed evidence contains gold & 64.0 \\
\bottomrule
\end{tabular}
\caption{Stage-wise loss analysis for the current best configuration. Packing is not the bottleneck (R3=0), while base recall and reranking remain the main failure points.}
\label{tab:stage_loss}
\end{table}

\subsection{Matched-Interface Retrieval-Stack Ablation}
We report here the strict matched-interface decomposition that holds the reader, prompt, packet interface, and evaluation split fixed while varying locality, the text-instantiated slice of \ours{}, visual assertions, and the conservative gate. The same pattern as in the main text reappears: the text-only slice of \ours{} supplies the main incremental gain, while visual assertions are slightly harmful in isolation but become modestly useful once locality-aware filtering is present.

\begin{table}[h]
\centering
\small
\setlength{\tabcolsep}{3pt}
\renewcommand{\arraystretch}{1.08}
\resizebox{\columnwidth}{!}{%
\begin{tabular}{@{}lccccccc@{}}
\toprule
Variant & Local & Text slice & Visual & Gate & EM & ANLS & ROUGE-L \\
\midrule
Page-only retrieval & -- & -- & -- & -- & 32.0 & 34.4 & 37.3 \\
Page-only + local prior & Yes & -- & -- & -- & 32.4 & 35.6 & 38.6 \\
PILAR (text-only slice) & -- & Yes & -- & -- & 33.2 & 36.1 & 38.9 \\
PILAR (text-only) + local & Yes & Yes & -- & -- & 33.0 & 36.1 & 39.0 \\
PILAR (text+visual slice) & -- & Yes & Yes & -- & 32.8 & 35.6 & 38.2 \\
PILAR (text+visual) + local & Yes & Yes & Yes & -- & 33.4 & 36.1 & 38.9 \\
Final (+ gate) & Yes & Yes & Yes & Yes & \textbf{33.6} & \textbf{36.3} & \textbf{39.1} \\
\bottomrule
\end{tabular}}
\caption{Strict matched-interface ablation (Naive RAG, M3DocVQA). All rows share the same reader, prompt, packet budget, and evaluation split. The table separates page-only retrieval, page-boundary locality, the text-only slice of \ours{}, visual assertions, and the conservative expansion gate under a single matched interface. The key comparison is between \emph{PILAR (text-only) + local} and \emph{PILAR (text+visual) + local}, which isolates the incremental contribution of visual assertions under fixed locality priors and packetization. The text-only slice of \ours{} provides the largest single gain (+1.2 EM over page-only), while visual assertions are slightly harmful in isolation but recover a +0.4 EM increment once locality-aware filtering is applied.}
\label{tab:matched_ablation}
\end{table}

A key empirical insight is that the current final ordering is \emph{answer-centric} rather than gold-page-centric. Gold-page MRR decreases slightly after local/graph reranking, but end-to-end EM improves. The local and graph-derived priors therefore act as evidence-ranking signals that surface useful non-gold pages, not merely as mechanisms for promoting the gold page itself. The strict ablation makes the interaction more precise: the text-instantiated slice of \ours{} is the main incremental component, while visual assertions are beneficial only once locality-aware filtering is present.

\subsection{Reader-Family Comparison}
We additionally compare the default 8B instruction-tuned VLM reader against a larger 14B text-only model under the same gold-page oracle protocol (raw gold-page text concatenation, same token budget, same evaluation). Table~\ref{tab:reader_family} shows that parameter count alone does not yield a stronger reader for this task: the 14B text-only model is worse on retrieved context and fails to exceed the 8B VLM under the gold-page oracle. In other words, a stronger reader for this task is not simply a bigger text-only model; model family and instruction tuning matter more than raw parameter count.

\begin{table}[h]
\centering
\small
\setlength{\tabcolsep}{3pt}
\renewcommand{\arraystretch}{1.08}
\resizebox{\columnwidth}{!}{%
\begin{tabular}{@{}llccc@{}}
\toprule
Reader & Context & All EM & Text EM & Vis.\ EM \\
\midrule
Qwen3-VL-8B & Retrieved final packet & 33.6 & 40.0 & 17.1 \\
Qwen3-14B & Retrieved final packet & 31.6 & 38.6 & 13.6 \\
Qwen3-VL-8B & Gold-page text only & 42.4 & 48.6 & 26.4 \\
Qwen3-14B & Gold-page text only & 42.0 & 47.8 & 27.1 \\
\bottomrule
\end{tabular}}
\caption{Reader-family comparison under a gold-page oracle protocol. A 14B text-only model does not outperform the 8B instruction-tuned VLM, either on retrieved packets or under the gold-page oracle.}
\label{tab:reader_family}
\end{table}

\subsection{Query-Type and Modality Oracle Analysis}
\label{sec:source_breakdown}

Table~\ref{tab:source_breakdown} stratifies \ours{} performance by query phenotype: \emph{text-grounded} questions (whose answer is typically recoverable from OCR text) versus \emph{visual-classified} questions (flagged by visual/appearance/layout cues in the query). Classification is performed with a keyword/pattern heuristic; importantly, a later audit shows that most of the visual-classified subset is still answerable from OCR text. Adding gold-page images to gold-page text does not improve---and slightly hurts---visual-classified EM (26.4 $\rightarrow$ 25.7), suggesting that at the current 8B scale the reader treats raw page images as noise rather than as complementary evidence.

\begin{table}[h]
\centering
\scriptsize
\setlength{\tabcolsep}{3pt}
\renewcommand{\arraystretch}{1.1}
\begin{tabular}{@{}lcccc@{}}
\toprule
& \multicolumn{2}{c}{EM} & \multicolumn{2}{c}{ANLS} \\
\cmidrule(lr){2-3}\cmidrule(lr){4-5}
Condition & Text & Visual & Text & Visual \\
& {\scriptsize(n=1,757)} & {\scriptsize(n=684)} & {\scriptsize(n=1,757)} & {\scriptsize(n=684)} \\
\midrule
\ours{} (retrieved) & 40.0 & 17.1 & 43.4 & 18.1 \\
\ours{} (gold text) & 48.6 & 26.4 & 51.7 & 26.9 \\
\ours{} (gold image) & --- & 18.6 & --- & 19.1 \\
\ours{} (gold image+text) & --- & 25.7 & --- & 26.7 \\
\midrule
$\Delta$ (gold text $-$ retrieved) & \up{+8.6} & \up{+9.3} & \up{+8.3} & \up{+8.8} \\
\bottomrule
\end{tabular}
\caption{\textbf{Query-type and modality oracle analysis} (Naive RAG, M3DocVQA). Text-grounded vs. visual-classified query performance under retrieved context, gold-page text, gold-page image, and gold-page image+text. The retrieval gap ($\Delta$) is similar for both subsets, but the visual-classified subset starts much lower.}
\label{tab:source_breakdown}
\end{table}

\subsection{Visual Answer-Source Audit and Modality Oracle}
To understand the hard visual-classified subset, we audit whether the answer is actually absent from OCR text or whether the reader simply fails to extract it. Table~\ref{tab:visual_audit} shows that approximately 74\% of visual-classified questions still have the answer in OCR text, while about 48\% are reader failures despite answer presence. This means that the visual-classified subset is hard, but not primarily because it requires image-only reasoning; under our audit criteria, the vast majority of these cases have an OCR-recoverable answer.

\begin{table}[h]
\centering
\small
\setlength{\tabcolsep}{4.5pt}
\renewcommand{\arraystretch}{1.08}
\begin{tabular}{lc}
\toprule
Category & \% \\
\midrule
Answer in OCR, reader correct & 26.4 \\
Answer in OCR, reader failed & 47.9 \\
No gold page in index & 25.7 \\
Answer absent from OCR & 0.0 \\
\bottomrule
\end{tabular}
\caption{Answer-source audit on the visual-classified subset. Most are still OCR-recoverable; the largest error source is reader failure despite answer presence in OCR.}
\label{tab:visual_audit}
\end{table}

We further compare text-only and image-based gold-page oracles on this subset. Table~\ref{tab:visual_oracle_modalities} shows that text-only gold pages outperform image-only gold pages, and that adding images to text does not improve over text alone. Under the current 8B VLM, the reader relies much more on OCR text than on raw page images.

\begin{table}[h]
\centering
\small
\setlength{\tabcolsep}{4.5pt}
\renewcommand{\arraystretch}{1.08}
\begin{tabular}{lccc}
\toprule
Input modality & EM & ANLS & ROUGE-L \\
\midrule
Retrieved final context & 17.1 & 18.1 & 20.4 \\
Gold page text only & 26.4 & 26.9 & 29.1 \\
Gold page image only & 18.6 & 19.1 & 21.1 \\
Gold page image + text & 25.7 & 26.7 & 29.6 \\
\bottomrule
\end{tabular}
\caption{Visual-classified subset modality oracle. Raw page images do not outperform OCR text, and image+text is not better than text alone at 8B scale.}
\label{tab:visual_oracle_modalities}
\end{table}

\subsection{Query-Attribute Analysis}
We reanalyze the M3DocVQA evaluation set by hop count, coarse question family, modality, and number of supporting documents to localize where the \ours{} retrieval stack helps most. The condensed summary now appears in main-text Table~\ref{tab:attribute_summary}. In short, the gains concentrate on compositional, comparison, two-document, and multi-modal subsets. The sampled 3-hop bucket shows the largest observed improvement, but because it is relatively small we treat it as suggestive rather than definitive.

\subsection{Frames as a Support-Chain Stress Test}
\label{sec:frames_stress}

We use Frames as a harder ODQA stress test for long support-chain recovery. In our hop-count analysis of the full Frames evaluation split, 770/824 questions (93.5\%) lie in the 2+hop regime, so we focus the detailed analysis on 2-hop, 3-hop, and 4+hop buckets rather than treating Frames as a precision-dominated single-hop benchmark. The shared cross-agent comparison in Table~\ref{tab:main_results} reports full-Frames results under the common retrieval configuration used elsewhere in the paper. Separately, a benchmark-matched full-set recomputation shows that the M3DocVQA-tuned single-shot setting is not appropriate for Frames: the adjacent-page bonus acts as noise, while a modestly broader pool is mildly helpful. Under the Frames-matched setting (no adjacent bonus, multiplier 4), \ours{} becomes the strongest directly comparable backend on the full Frames split, although the margins remain modest. We treat this Frames-matched result as a robustness analysis on Frames rather than as a held-out hyperparameter-selection outcome.

\begin{table}[h]
\centering
\footnotesize
\setlength{\tabcolsep}{3pt}
\renewcommand{\arraystretch}{1.08}
\begin{tabular}{lccc}
\toprule
Backend & EM & ANLS & ROUGE-L \\
\midrule
\ours{} (Frames-matched) & \textbf{14.2} & \textbf{17.8} & \textbf{21.3} \\
Flat chunk & 13.4 & 17.1 & 20.2 \\
\ours{} w/o entity exp. & 13.2 & 16.8 & 20.1 \\
MS GraphRAG & 13.0 & 16.6 & 19.7 \\
LightRAG & 12.4 & 15.7 & 18.6 \\
MultiDocFusion & 12.4 & 15.8 & 18.4 \\
\bottomrule
\end{tabular}
\caption{Full-set Frames comparison under the Frames-matched setting (\texttt{initial\_multiplier}=4, \texttt{adjacent\_page\_bonus}=0). \ours{} is the strongest directly comparable backend overall, but the margin over flat retrieval remains modest (+0.8 EM, +0.7 ANLS).}
\label{tab:frames_fullset}
\end{table}

To understand why the gain remains small, we compare retrieval tuning and oracle variants on the 2+hop subset. Removing the local prior alone improves \ours{} by +1.7 EM and +2.0 ANLS over the original Frames baseline, while widening the pool from multiplier 2 to 4 adds only a further +0.4 EM. This shows that the main fix is not broader exploration by itself, but removing a locality prior that is mismatched to support-chain retrieval.

\begin{table}[h]
\centering
\footnotesize
\setlength{\tabcolsep}{3pt}
\renewcommand{\arraystretch}{1.08}
\begin{tabular}{@{}lcc@{}}
\toprule
Condition & 2+hop EM & 2+hop ANLS \\
\midrule
F0: original baseline & 9.7 & 13.2 \\
F1: no adj.\ bonus & 11.4 & 15.2 \\
Frames-matched & 11.8 & 15.4 \\
\midrule
O1: gold answer page only & 3.4 & 4.7 \\
O2: all gold support pages & 17.7 & 21.7 \\
\bottomrule
\end{tabular}
\caption{Frames 2+hop stress-test summary. The answer page alone is insufficient; the gold support-chain oracle is far stronger than the gold answer-page oracle, which shows that support-chain completeness matters more than answer-page recovery alone. Yet even the support-chain oracle remains low, indicating a substantial reader/reasoning bottleneck.}
\label{tab:frames_stress_summary}
\end{table}

Coverage diagnostics explain why the end-to-end margin remains modest. Under the current objective, \ours{} achieves only slightly higher support-chain coverage than flat retrieval (e.g., all-support-docs@10 of 53.2\% vs. 52.1\%), while retrieving more diverse but also more dispersed evidence. We therefore interpret Frames as evidence that the current \ours{} objective is better aligned with evidence aggregation than with strict support-chain reconstruction. This does not negate the M3DocVQA result; rather, it shows that benchmark-specific retrieval objectives matter, and that locality-aware priors can switch from helpful to harmful depending on whether the benchmark emphasizes page continuity or support-chain completeness.

% Additional color definitions for qualitative analysis
\FloatBarrier
\clearpage
\section{Part II: PILAR-Specific Analysis --- Retrieval Trace and Comparison}
\label{sec:retrieval_trace}

\subsection{Query-Driven Retrieval Trace}

\begin{figure}[h]
\centering
\resizebox{\columnwidth}{!}{%
\begin{tikzpicture}[
  font=\scriptsize,
  >=Latex,
  stage/.style={draw, rounded corners=3pt, fill=systemcolor, thick, align=center, minimum height=8mm, minimum width=28mm, inner sep=2pt},
  entity/.style={draw, rounded corners=2pt, fill=blue!8, align=center, minimum height=6mm, inner sep=2pt},
  assertion/.style={draw, rounded corners=2pt, fill=yellow!10, align=center, minimum height=5.5mm, inner sep=1.5pt},
  support/.style={draw, rounded corners=1.5pt, fill=green!8, align=center, minimum height=5mm, inner sep=1.5pt, font=\tiny},
  packet/.style={draw, rounded corners=3pt, fill=oursrow, thick, align=center, minimum height=8mm, inner sep=2pt},
  edgelbl/.style={font=\tiny, fill=white, inner sep=0.5pt},
  fade/.style={opacity=0.3},
]

% Query
\node[stage, fill=white, draw=black!60] (q) at (0, 0) {\textbf{Query:} ``What is the\\max RPM of Engine-X?''};

% Stage 1: Base shortlist
\node[stage] (base) at (0, -1.4) {\textbf{1. Base Shortlist}\\{\tiny BM25+dense $\rightarrow$ top-40 pages}};
\draw[->, thick] (q) -- (base);

% Stage 2: Seed matching
\node[stage] (seed) at (0, -2.8) {\textbf{2. Seed Match}\\{\tiny query $\rightarrow$ entity/assertion fields}};
\draw[->, thick] (base) -- (seed);

% Seed results
\node[entity] (e1) at (-2.2, -4.2) {\textbf{Engine-X}\\[-1pt]{\tiny seed entity}};
\node[assertion] (a1) at (2.2, -4.2) {max\_rpm = 5000\\[-1pt]{\tiny seed assertion}};
\draw[->, thick, blue!50] (seed) -- (e1);
\draw[->, thick, yellow!70!black] (seed) -- (a1);

% Stage 3: Classify + Expand
\node[stage, minimum width=55mm] (expand) at (0, -5.6) {\textbf{3. Classify (E)} $\rightarrow$ \textbf{Graph Expand}\\{\tiny follow entity $\rightarrow$ assertion $\rightarrow$ support edges}};
\draw[->, thick] (e1) -- (expand);
\draw[->, thick] (a1) -- (expand);

% Expanded neighborhood
\node[assertion, fade] (a2) at (-2.8, -7.0) {exception:\\4500 if cond\_C};
\node[support] (s1) at (-0.8, -7.0) {Table 3, row 2\\{\tiny primary}};
\node[support] (s2) at (1.2, -7.0) {\S2.1 para 3\\{\tiny contextual}};
\node[support, fill=orange!10] (s3) at (3.2, -7.0) {Fig 5 crop\\{\tiny visual}};
\draw[->, thick, densely dotted] (expand) -- (a2);
\draw[->, thick] (expand) -- (s1);
\draw[->, thick] (expand) -- (s2);
\draw[->, thick, orange!60] (expand) -- (s3);

% Stage 4: evidence packet
\node[packet, minimum width=60mm] (pack) at (0, -8.4) {\textbf{4. Evidence Packet} (within token budget $B_{\text{tok}}$)\\{\tiny page context + support metadata + visual crops}};
\draw[->, thick] (s1) -- (pack);
\draw[->, thick] (s2) -- (pack);
\draw[->, thick, orange!60] (s3) -- (pack);

% Arrow to reader
\node[stage, fill=gray!10, draw=gray!50] (reader) at (0, -9.6) {\textbf{Reader} (Qwen3-VL-8B)};
\draw[->, thick] (pack) -- (reader);

\end{tikzpicture}}%
\caption{Query-driven retrieval trace in \ours{}. A query enters the base shortlist, seeds are matched against entity/assertion fields, the query classifier selects an expansion profile, and graph traversal collects a support neighborhood. The packet builder assembles the final token-budgeted packet for the reader.}
\label{fig:retrieval_trace}
\end{figure}
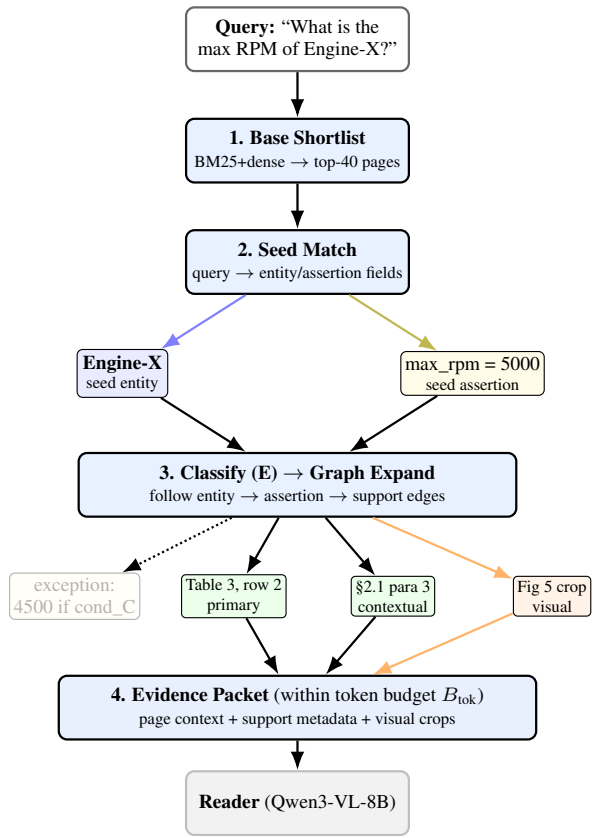

Figure~\ref{fig:retrieval_trace} illustrates how \ours{} processes a single query through the seed $\rightarrow$ expand $\rightarrow$ pack pipeline. Starting from the base page shortlist, the system matches query terms against entity and assertion fields to find seed nodes. The query classifier selects an expansion profile, and graph traversal follows entity--assertion--support edges to collect a neighborhood. The packet builder then assembles the final packet under the token budget.

\subsection{Flat Retrieval vs.\ \ours{}}

\begin{figure*}[h]
\centering
\resizebox{\textwidth}{!}{%
\begin{tikzpicture}[
  font=\scriptsize,
  >=Latex,
  chunk/.style={draw, rounded corners=2pt, fill=gray!8, align=center, minimum height=7mm, inner sep=2pt},
  entity/.style={draw, rounded corners=2pt, fill=blue!8, thick, align=center, minimum height=7mm, inner sep=2pt},
  assertion/.style={draw, rounded corners=2pt, fill=yellow!10, align=center, minimum height=6.5mm, inner sep=2pt},
  support/.style={draw, rounded corners=1.5pt, fill=green!8, align=center, minimum height=6mm, inner sep=1.5pt, font=\tiny},
  packet/.style={draw, rounded corners=3pt, thick, align=center, minimum height=8mm, inner sep=3pt},
  miss/.style={draw, rounded corners=2pt, fill=red!8, align=center, minimum height=6mm, inner sep=2pt, densely dashed, draw=red!40},
  hit/.style={draw, rounded corners=2pt, fill=green!12, align=center, minimum height=6mm, inner sep=2pt, thick, draw=green!50!black},
  edgelbl/.style={font=\tiny, fill=white, inner sep=0.5pt},
  titlestyle/.style={font=\small\bfseries},
]

% ============ LEFT: Flat Retrieval ============
\node[titlestyle] at (-5, 1.0) {Flat Chunk Retrieval};

\node[chunk] (q1) at (-5, 0) {Query: ``Max RPM of Engine-X\\and any exceptions?''};

% Retrieved chunk
\node[hit] (c1) at (-6.5, -1.6) {Chunk: ``Engine-X\\max RPM = 5000''\\{\tiny Doc 1, \S2.1}};
\node[hit] (c2) at (-3.5, -1.6) {Chunk: ``Engine-X\\specs overview''\\{\tiny Doc 1, \S2.0}};

% Missed
\node[miss] (m1) at (-6.5, -3.4) {{\tiny\color{red!60} MISSED}\\Exception: 4500 rpm\\if condition\_C\\{\tiny Doc 2, Fig 5 caption}};
\node[miss] (m2) at (-3.5, -3.4) {{\tiny\color{red!60} MISSED}\\Wear pattern chart\\{\tiny Doc 2, Fig 5 crop}};

\draw[->, thick] (q1) -- (c1);
\draw[->, thick] (q1) -- (c2);
\draw[->, thick, red!40, densely dashed] (q1) -- (m1) node[midway, left, font=\tiny, text=red!60] {no bridge};
\draw[->, thick, red!40, densely dashed] (q1) -- (m2) node[midway, right, font=\tiny, text=red!60] {no bridge};

% Flat packet
\node[packet, fill=gray!8, minimum width=40mm] (fp) at (-5, -5.0) {\textbf{Flat Packet}\\{\tiny Doc 1 chunks only}\\{\tiny missing exception + visual}};

\draw[->, thick] (c1) -- (fp);
\draw[->, thick] (c2) -- (fp);

% ============ RIGHT: PILAR ============
\node[titlestyle] at (5, 1.0) {\ours{} Retrieval};

\node[chunk, fill=white] (q2) at (5, 0) {Query: ``Max RPM of Engine-X\\and any exceptions?''};

% Entities
\node[entity] (e1) at (3.2, -1.4) {\textbf{Engine-X}\\{\tiny Doc 1}};
\node[entity] (e2) at (6.8, -1.4) {\textbf{Engine-X}\\{\tiny Doc 2}};
\draw[<->, thick, blue!60, densely dashed] (e1) -- node[edgelbl, above]{\texttt{same\_as}} (e2);

\draw[->, thick] (q2) -- (e1);
\draw[->, thick] (q2) -- (e2);

% Assertions + supports from Doc 1
\node[assertion] (a1) at (2.2, -2.8) {max\_rpm = 5000\\{\tiny text-derived}};
\node[support] (s1) at (2.2, -4.0) {Table 3, row 2\\{\tiny primary support}};
\draw[->, thick, blue!50] (e1) -- (a1);
\draw[->, thick, yellow!70!black] (a1) -- (s1);

% Assertions + supports from Doc 2
\node[assertion] (a2) at (5.8, -2.8) {exception: 4500\\{\tiny if condition\_C}};
\node[assertion] (a3) at (8.0, -2.8) {wear = abnormal\\{\tiny VLM assertion}};
\node[support] (s2) at (5.8, -4.0) {Fig 5 caption\\{\tiny text support}};
\node[support, fill=orange!10] (s3) at (8.0, -4.0) {Fig 5 crop\\{\tiny visual support}};
\draw[->, thick, blue!50] (e2) -- (a2);
\draw[->, thick, blue!50] (e2) -- (a3);
\draw[->, thick, yellow!70!black] (a2) -- (s2);
\draw[->, thick, orange!60, densely dotted] (a3) -- (s3);

% PILAR packet
\node[packet, fill=oursrow, minimum width=65mm] (vp) at (5, -5.4) {\textbf{Evidence Packet}\\{\tiny Doc 1: spec table + text \quad Doc 2: exception + visual crop}\\{\tiny cross-document evidence assembled via entity bridge}};
\draw[->, thick] (s1) -- (vp);
\draw[->, thick] (s2) -- (vp);
\draw[->, thick, orange!60] (s3) -- (vp);

\end{tikzpicture}}%
\caption{Flat retrieval vs.\ \ours{} on the same cross-document query. \textbf{Left:} Flat chunk retrieval returns top-scoring chunks from Doc~1 but misses the cross-document exception and visual evidence in Doc~2 because no entity bridge exists. \textbf{Right:} \ours{} links Engine-X across documents via a \texttt{same\_as} edge, retrieves both text-derived and VLM-derived assertions, and assembles a complete packet that includes the exception qualifier and the visual wear-pattern evidence.}
\label{fig:flat_vs_ours}
\end{figure*}
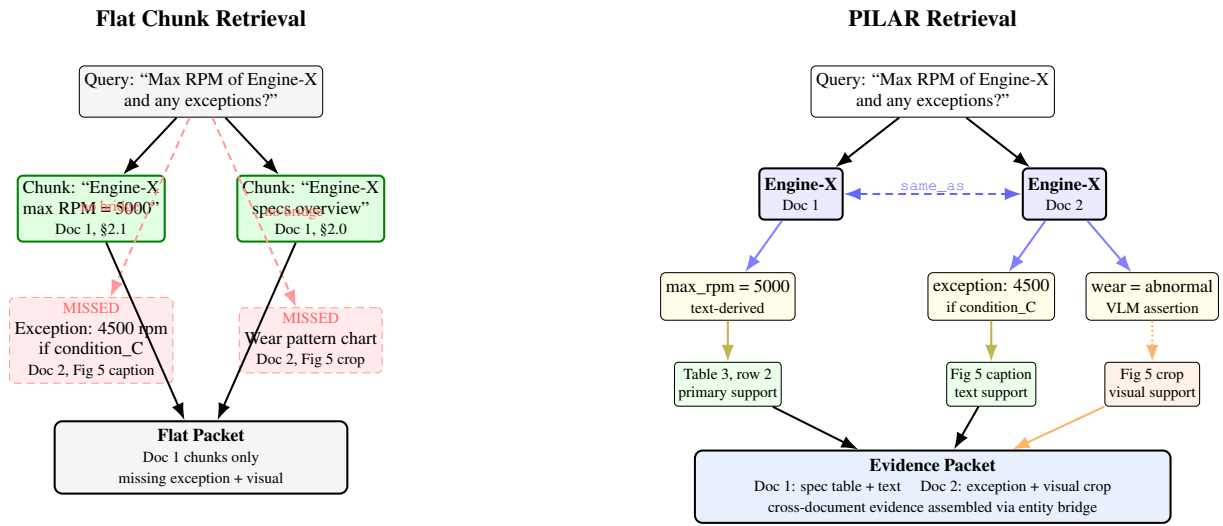

Figure~\ref{fig:flat_vs_ours} contrasts flat chunk retrieval with \ours{} on the same cross-document query. Flat retrieval returns the highest-scoring page from one document but misses the cross-document qualifier and visual evidence. \ours{} uses the entity bridge to traverse to the second document and includes both the text-derived exception and the VLM-derived visual assertion in the final packet.

\FloatBarrier
\clearpage
\section{Part II: PILAR-Specific Analysis --- Runtime and Scalability Analysis}
\label{sec:runtime_supp}

We analyze the computational cost of \ours{} under a realistic industrial setting.
Measurements are conducted on a single NVIDIA H100 (80GB) GPU with 144 DPI rendering and high-precision OCR settings.

\subsection{Offline Indexing Cost}
Table~\ref{tab:supp_runtime} breaks down the processing runtime by document length.

\begin{table}[!t]
\centering
\scriptsize
\renewcommand{\arraystretch}{1.2}
\setlength{\tabcolsep}{2.0pt}

\resizebox{\columnwidth}{!}{%
\begin{tabular}{@{}l ccc c@{}}
\toprule
\multirow{2}{*}{Stage} & \multicolumn{3}{c}{Latency (s)} & \multirow{2}{*}{Complexity} \\
\cmidrule(lr){2-4}
& 5 pages & 10 pages & 20 pages & \\
\midrule
\multicolumn{5}{l}{\textit{Visual Analysis (High-Res)}} \\
\quad Layout Detection & 5.2 & 10.5 & 21.2 & $O(P)$ \\
\quad OCR (Dense Text) & 24.5 & 48.2 & 98.4 & $O(P \times T)$ \\
\quad Visual Embedding & 12.5 & 25.8 & 52.6 & $O(N_{\text{img}})$ \\
\midrule
\multicolumn{5}{l}{\textit{Structure \& Graph Construction}} \\
\quad Structure Parse & 0.5 & 1.0 & 2.1 & $O(N_{\text{blk}})$ \\
\quad Graph Build & 0.2 & 0.4 & 0.9 & $O(N + E)$ \\
\midrule
\multicolumn{5}{l}{\textit{Assertion Induction}} \\
\quad Text-only LLM (batch) & 1--4 & 3--8 & 6--16 & $O(N_{\text{text}})$ \\
\quad VLM visual extraction (batch) & 3--8 & 6--16 & 12--32 & $O(N_{\text{visual}})$ \\
\quad Self-verif.\ + span alignment & 0.2 & 0.4 & 0.8 & $O(N_{\text{induced}})$ \\
\quad Predicate normalization & 0.1 & 0.1 & 0.2 & $O(N_{\text{pred}})$ \\
\midrule
Total Indexing Time & 46.6--51.6 & 93.2--103.2 & 189.8--209.8 & Linear w.r.t Pages \\
\bottomrule
\end{tabular}%
}
\caption{
Offline indexing runtime and scalability. The \textit{Assertion Induction} block reports the cost of the modality-based extraction pipeline. VLM latency is shown as a range because the number of visual elements varies by document. Across the measured setup, visual analysis remains the dominant cost, while the assertion induction stages add bounded offline overhead.
}
\label{tab:supp_runtime}
\end{table}

\paragraph{Findings and Cost Justification.}
Visual analysis (layout, OCR, embeddings) remains the dominant component of the offline cost under the setup reported in Table~\ref{tab:supp_runtime}. The additional cost of assertion induction is concentrated in the VLM path for visual elements, while the text-only LLM path and post-processing (verification, predicate normalization) are comparatively small. Section-level batching keeps model invocations practical, and the assertion induction stages scale roughly linearly with document size in the measured regime.

\FloatBarrier

% Bibliography file must be supplied with the submission package.
\end{document}